\documentclass[8.5pt,twoside,twocolumn]{article}
\usepackage[super,sort&compress,comma]{natbib}
\usepackage{mhchem}
\usepackage{times,mathptmx}
\usepackage{sectsty}
\usepackage{balance}

\usepackage{graphicx} 
\usepackage{lastpage}
\usepackage[format=plain,justification=raggedright,singlelinecheck=false,font=small,labelfont=bf,labelsep=space]{caption}
\usepackage{fancyhdr}
\begin{document}

\thispagestyle{plain}
\fancypagestyle{plain}{
\fancyhead[L]{\includegraphics[height=8pt]{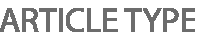}}
\fancyhead[C]{\hspace{-1cm}\includegraphics[height=20pt]{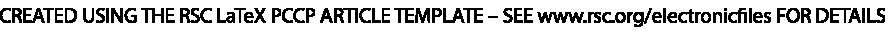}}
\fancyhead[R]{\includegraphics[height=10pt]{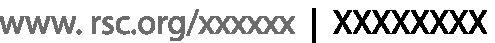}\vspace{-0.2cm}}
\renewcommand{\headrulewidth}{1pt}}
\renewcommand{\thefootnote}{\fnsymbol{footnote}}
\renewcommand\footnoterule{\vspace*{1pt}%
\hrule width 3.4in height 0.4pt \vspace*{5pt}}
\setcounter{secnumdepth}{5}

\newcommand{\meanv}[1]{\left\langle#1\right\rangle}
\newcommand{\R}{\mathbb{R}}
\newcommand{\1}{\mathbf{1}}
\newcommand{\C}{\mathbb{C}}
\newcommand{\I}{\mathrm{i}}
\newcommand{\E}{\mathbb{E}}
\newcommand{\D}{\mathrm{d}}
\newcommand{\N}{\mathbb{N}}
\newcommand{\Z}{\mathbb{Z}}
\newcommand{\Q}{\mathbb{Q}}
\newcommand{\T}{\mathbb{T}}

\newcommand{\hi}{\mathcal{H}}
\newcommand{\B}{\mathcal{B}}
\newcommand{\U}{\mathcal{U}}
\newcommand{\F}{\mathcal{F}}
\newcommand{\M}{\mathcal{M}}
\newcommand{\W}{\mathcal{W}}
\newcommand{\J}{\mathcal{J}}
\newcommand{\Do}{\mathcal{D}}
\newcommand{\Sch}{\mathcal{S}}
\newcommand{\A}{\mathcal{A}}
\newcommand{\Or}{\mathcal{O}}
\newcommand{\Lap}{\Delta}
\newcommand{\Li}{\mathcal{L}}
\newcommand{\epsi}{\varepsilon}
\newcommand{\ph}{\varphi}
\newcommand{\ix}{\index}
\newcommand{\indexd}{\in\{ 1,\ldots,d \}}
\newcommand{\ssp}{\left\langle }

\newcommand{\be}{\begin{equation}}
\newcommand{\ee}{\end{equation}}

\newcommand{\OO}[1]{O \left(\frac{1}{#1}\right)}

\def\Var{\operatorname{Var}}

\def\s{\sigma}
\def\f{\varph_i}
\def\b{\beta}
\def\g{\gamma}
\def\e{\eta}
\def\d{\delta}
\def\l{\lambda}
\def\r{\rho}
\def\tr{\tilde\rho}
\def\Zn{ \left\langle Z^{n}\right\rangle}
\def\dm{\mathcal{D}m_\mu^a}
\def\sumab{\sum_{a,b}^{n}}
\def\dqr{\prod_{a,b}^n dq_{ab} dr_{ab}}
\def\dmnu{\mathcal{D}m_\nu^a}
\def\sumh{\sum_{\{h\}}}
\def\Be'{\beta_\mu^{'}}
\def\q{q_{ab}}
\def\<{\bigl\langle}
\def\>{\bigr\rangle}
\def\dq{\mathcal{D}q_{ab}}
\def\lan{\bigl\langle\!\bigl\langle}
\def\ran{\bigr\rangle\!\bigr\rangle}
\def\ll{\langle}
\def\rr{\rangle}

\makeatletter
\def\subsubsection{\@startsection{subsubsection}{3}{10pt}{-1.25ex plus -1ex minus -.1ex}{0ex plus 0ex}{\normalsize\bf}}
\def\paragraph{\@startsection{paragraph}{4}{10pt}{-1.25ex plus -1ex minus -.1ex}{0ex plus 0ex}{\normalsize\textit}}
\renewcommand\@biblabel[1]{#1}
\renewcommand\@makefntext[1]%
{\noindent\makebox[0pt][r]{\@thefnmark\,}#1}
\makeatother
\renewcommand{\figurename}{\small{Fig.}~}
\sectionfont{\large}
\subsectionfont{\normalsize}

\fancyfoot{}
\fancyfoot[LO,RE]{\vspace{-7pt}\includegraphics[height=9pt]{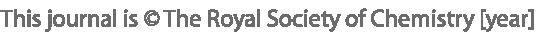}}
\fancyfoot[CO]{\vspace{-7.2pt}\hspace{12.2cm}\includegraphics{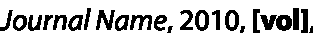}}
\fancyfoot[CE]{\vspace{-7.5pt}\hspace{-13.5cm}\includegraphics{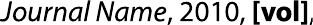}}
\fancyfoot[RO]{\footnotesize{\sffamily{1--\pageref{LastPage} ~\textbar  \hspace{2pt}\thepage}}}
\fancyfoot[LE]{\footnotesize{\sffamily{\thepage~\textbar\hspace{3.45cm} 1--\pageref{LastPage}}}}
\fancyhead{}
\renewcommand{\headrulewidth}{1pt}
\renewcommand{\footrulewidth}{1pt}
\setlength{\arrayrulewidth}{1pt}
\setlength{\columnsep}{6.5mm}
\setlength\bibsep{1pt}

\twocolumn[
  \begin{@twocolumnfalse}
\noindent\LARGE{\textbf{Cancer-driven dynamics of immune cells in a microfluidic environment}}
\vspace{0.6cm}

\noindent\large{\textbf{Elena Agliari,\textit{$^{a}$}  Elena Biselli, \textit{$^{a}$} Adele De Ninno, \textit{$^b$}  Giovanna Schiavoni, \textit{$^{c}$} Lucia Gabriele,\textit{$^{c}$}  Anna Gerardino, \textit{$^{c}$} Fabrizio Mattei, \textit{$^{c,\dag}$} Adriano Barra,\textit{$^{a,\dag}$} Luca Businaro \textit{$^{b,\dag}$}   }}\vspace{0.5cm}

\noindent\textit{\small{\textbf{Received Xth XXXXXXXXXX 20XX, Accepted Xth XXXXXXXXX 20XX\newline
First published on the web Xth XXXXXXXXXX 200X}}}

\noindent \textbf{\small{DOI: 10.1039/b000000x}}
\vspace{0.6cm}

\noindent \normalsize{
{\em Background} Over the past decade, the dynamical interactions occurring between immune system cells and tumors have been object of intense studies, even at multidisciplinary level. In particular, recent investigations have aimed to apply mathematical models, such as the stochastic process theory, in order to predict the behavioral parameters of the biological phenomena. Scope of the present work is to infer the migratory ability of leukocytes by the random process theory in order to distinguish the spontaneous re-organization of immune cells against cancer. For this purpose, splenocytes from immunodeficient mice, namely selectively lacking the transcription factor IRF-8 (IRF-8 KO), or from immunocompetent animals (wild-type; WT), were allowed to interact, alternatively, with murine B16.F10 melanoma cells in an ad hoc microfluidic environment developed on a LabOnChip technology. In this setting, only WT splenocytes were able to interact with melanoma cells and to coordinate a response against the tumor cells through physical interaction. Conversely, IRF-8 KO immune cells exhibited poor dynamical reactivity towards the neoplastic cells.
\newline
{\em Results} We collected and analyzed data on the motility of the cells and found, with remarkable accuracy, that the IRF-8 KO cells performed pure uncorrelated random walks. Instead, WT splenocytes were able to make singular drifted random walks that, averaged over the ensemble of cells, collapsed on a straight ballistic motion for the system as a whole, hence giving rise to a coordinate response. At a finer level of investigation, we found that IRF-8 KO splenocytes moved rather uniformly since their step lengths were exponentially distributed, while WT counterparts displayed a qualitatively broader motion, as their step lengths along the direction of the melanoma were log-normally distributed.
\newline
{\em Conclusions} Summarizing, our findings on single WT leukocyte dynamics reveal very broad quasi-random flows, while coarse-graining on the system as a whole, this moves quite as a rigid body; further, this kind of investigation suggests that Cell-on-Chip tools allowing under-microscope-like experiments may imply a cascade of promising quantitative techniques, whose application here works as a test case of many possible.}
\vspace{0.5cm}
 \end{@twocolumnfalse}
  ]

\footnotetext{\dag~These authors share equal credit for senior authorship.}


\footnotetext{\textit{$^{a}$Dipartimento di Fisica, Sapienza Universita' di Roma.}}
\footnotetext{\textit{$^{b}$CNR, Istituto di Nanofotonica, Roma.}}
\footnotetext{\textit{$^{c}$Istituto Superiore di Sanita', Roma.}}



\section{Introduction}
The quest for the development of theoretical frames able to describe biological systems has been a {\em leitmotif} in  the work of many physicist and mathematicians since when biologists unveiled the complexity of the phenomena at the base of our existence. The questions to be answered and some of the tools that people have used to answer them, where already present in the book ``What is life'' written by Erwin Schroedinger in 1944. In particular the tools made available by stochastic processes and statistical mechanics, which were in great ferment already in those years, where recognized as fundamental, given the {\em  not-fully-deterministic interaction of many body} present in every live organism.
Quite obviously, the development of theoretical frames able to describe the behavior of interacting cells,  or cell populations, is deeply linked with the possibility to obtain measurements of the parameters and variables describing the system under observation.
\newline
To increase the latter, in these last years we assisted to a great advancement of the methodologies employed to represent the multifaceted phenomena behind the biological and molecular events of the cell at the empirical level. Imaging and real time representation of cellular events occurring in the actually existing and recognized biology systems have always been regarded as a fascinating set of tool to follow in detail these phenomena.
\newline
One of the main scientific fields actively working in this context are represented by investigations regarding the fight against cancer. The literature debating on this vast topic is largely increasing overtime \cite{1,2}.
\newline
This very broad literature witnesses on how the research regarding this field of investigation is moving to. We assisted to a parallel advancement of the techniques used to explore the cellular and molecular events, and to the development of methods aimed at highlighting the importance of considering the system as a whole, hence with all its constituent properly interacting. While the firsts tackle the basis of cancer progression mechanisms focusing on details of a single subject (cell or molecule), the seconds are becoming a major topic, involved as a necessary step beyond reductionism limitations \cite{quake, bialek2,ABMG2011,AABCT2013b}.
\newline
To strength this perspective, complex experimental systems such as confocal microscopy \cite{wilson}, two photon microscopy \cite{miller}, scanning electron microscopy \cite{goldstein} as well as transmission electron microscopy \cite{williams} are available to researchers, allowing to deeply investigate on several details of intracellular particles and cell-cell interactions. Nevertheless, great strides have been made over the past decade in this context, by the availability of innovative approaches to finely follow some key events occurring in cancer progression. In this regard, microfluidic systems have been proven valid and innovative platforms to approach on cancer cell interactions with drugs and chemotactic stimuli, and this gave a decisive gain in the dissection of the multiple mechanisms on how these compounds translate these signals in terms of migration \cite{3,4,5}.
\newline
Despite the gigantic advancement made by microscopy and molecular biology, it is still difficult to obtain such data from complex systems. Taking as example the immune system, quoting Kim and coworkers, it {\em operates according to a diverse, interconnected network of interactions, and the complexity of the network makes it difficult to understand experimentally. On one hand, in vitro experiments that examine a few or several cell types at a time often provide useful information about isolated immune interactions. However, these experiments also separate immune cells from the natural context of a larger biological network, potentially leading to non-physiological behavior. On the other hand, in vivo experiments observe phenomena in a physiological context, but are usually incapable of resolving the contributions of individual regulatory components} \cite{immune}.
In our view, the emerging solution to the limitations of the in vitro and in vivo experiments, dwells,  in the emerging field of the reconstitution of cellular microenvironment, relies on exploiting microfluidic chips and cell co-cultures. This approach, which we henceforth call Cells-On-Chip, has the great advantage of making realistic models \cite{Zervantonakis, Wlodkovitc}  of in vivo environment onto substrate perfectly compatible to modern microscopy tools and molecular biology methods. These devices therefore constitute the lacking bridge between biology and theoretical models as we and other groups demonstrated\cite{quake}.
\newline
In the theoretical counterpart, strongly supported by the former experimental information finally available, modelers coming from different disciplines (e.g., mathematics and theoretical physics) are starting to adapt their skills to the biological subject. Thus several techniques, such as maximum entropy principle \cite{bialek1}, disordered statistical mechanics \cite{AABCT2013b}, complex optimization \cite{enzo}, graph theory \cite{ton}, stochastic processes \cite{AABSVW2013} and dynamical systems \cite{Perelson-RevModPhys1997} are starting to concretely contribute to the field, and mathematical models have been recently employed to study distinct properties of different types of cancer \cite{22, 26}.
\newline
In this scenario, the interactions between cancer cells and immune cells within the tumor micro-environment attracted
 many researchers, and an increasing number of reports has been produced so far \cite{6, 7, 8, 9, 10, 11, 12, 13}. Nevertheless, the available literature provided poor experimental information in the context of live imaging on dynamic events and cellular interactions occurring during such a crosstalk. In this regard, we recently exploited microfluidic devices to investigate in real time on the mutual interactions between cancer and immune cells \cite{Businaro-LC2013}. We employed a co-colture microfluidic system, composed by a set of four interconnected microchannel arrays, to investigate on the crosstalk between mouse melanoma cells and immune cells. To do so, we took advantage of spleen cells from mice deficient for the transcription factor IRF-8 \cite{16, 17, 18, 19, 20, 21}, essential for the induction of competent immune responses, and compared several mutual migratory parameters to those observed in presence of WT splenocytes.

Thus, the aim of the present work is to properly frame the outcome of the differential cancer-driven dynamics of immune cells  emerged in on chip experiments, described in section two, into a stochastic theory: overall, our findings suggest that predictive mathematical models are well suitable for microfluidic co-colture systems and reveal a potential usefulness in further discovering of novel parameters to be correlated to migratory phenomena of cancer and immune cells.
Overall, our findings suggest that predictive mathematical models are well suitable for microfluidic co-colture systems and reveal a potential usefulness in further discovering of novel parameters to be correlated to migratory phenomena of cancer and immune cells.

\section{Experiment and data description}

The data analyzed in this paper were gathered from an experiment described in detail elsewhere \cite{Businaro-LC2013}. Major biological results of such experiment  will be summarized hereafter for reader convenience. The basic idea was to reproduce on chip the interactions between cells of the immune system with the tumor, mimicking as much as possible those occurring in vivo. To do so, we realized a microfluidic chip, shown in Figures $1,2$, which was basically divided in two different zones, one for the culture of immune cells (marked in red), and the second for the tumor (green chamber), which consisted of the murine metastatic melanoma cell line B16.F10.
\begin{figure}[h!]
\includegraphics[width=0.46\textwidth]{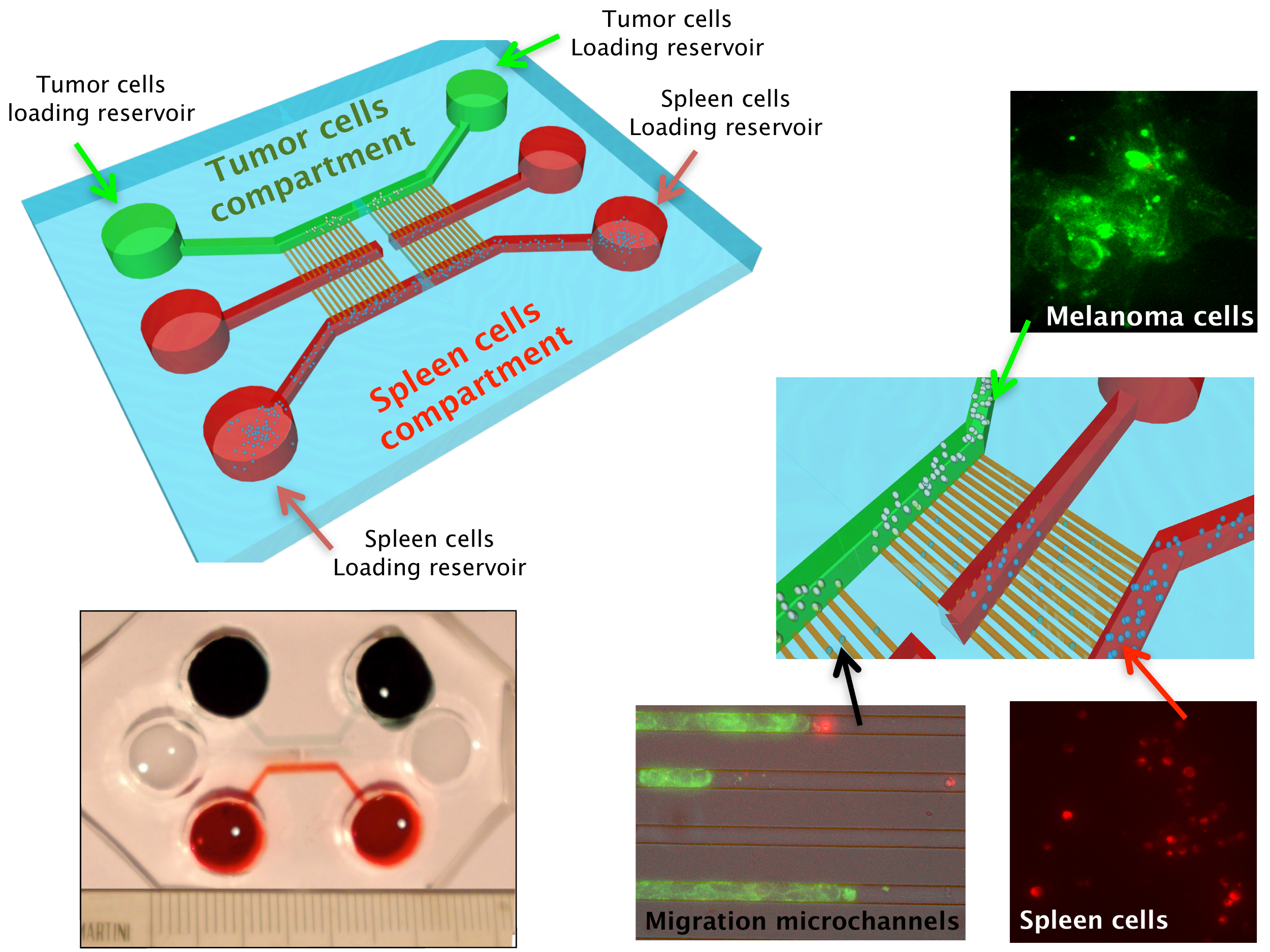}
\caption{Micro-fluidic co-culture immune-cancer system: design and fabrication.}
\label{fig:Settings}
\end{figure}
The two zones were connected by an array of microchannels (section $10 \times 10mm^2$) which allowed the migration of cells.
We used for the immune system a mouse spleen homogenate, which contains the whole pool of mature immune cell populations, ranging from T and B lymphocytes to phagocytes.  The experiments were carried out taking advantage of  two different sets of splenocytes, the first was from a wild type (WT) mouse, namely a healthy immune system, the second was from a mouse deficient for the transcription factor IRF-8, essential for the induction of competent immune responses. For the purposes of this paper, it means that we had a chip with a competent immune system facing the melanoma, while in a second chip we had a knocked-out (KO) immune system that we expected to be not responding to signals from the tumor cells.
\begin{figure}[h!]
\includegraphics[width=0.46\textwidth]{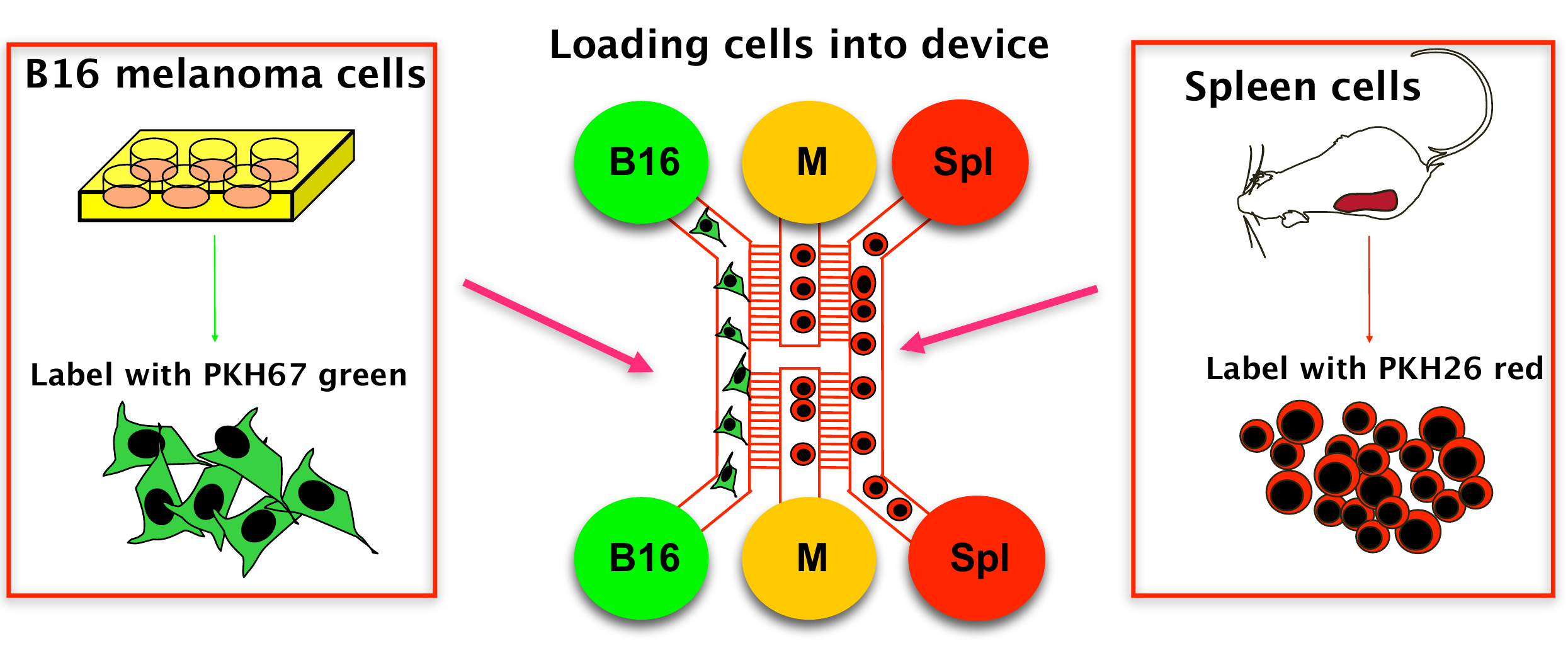}
\caption{Experimental design and methodologies: melanoma vs WT/KO immune cell interactions.}
\label{fig:Cartoon}
\end{figure}
In this experimental setting, both melanoma and immune cells could mutually migrate through the whole microfluidic system.The two systems were monitored by means of fluorescence microscopy up to 144 h and time-lapse recordings of the first 48 h of culture. For the B16-WT system we observed  a clear migration of the immune cells towards the melanoma and the formation of immune cell cluster around the B16 cells (see Figure 3). In the case of the B16-KO system, on the contrary, nor the KO cells show any response to the melanoma neither their trajectories were focused toward the insult (see Figure 4).
\begin{figure}[h!]
\includegraphics[width=0.46\textwidth]{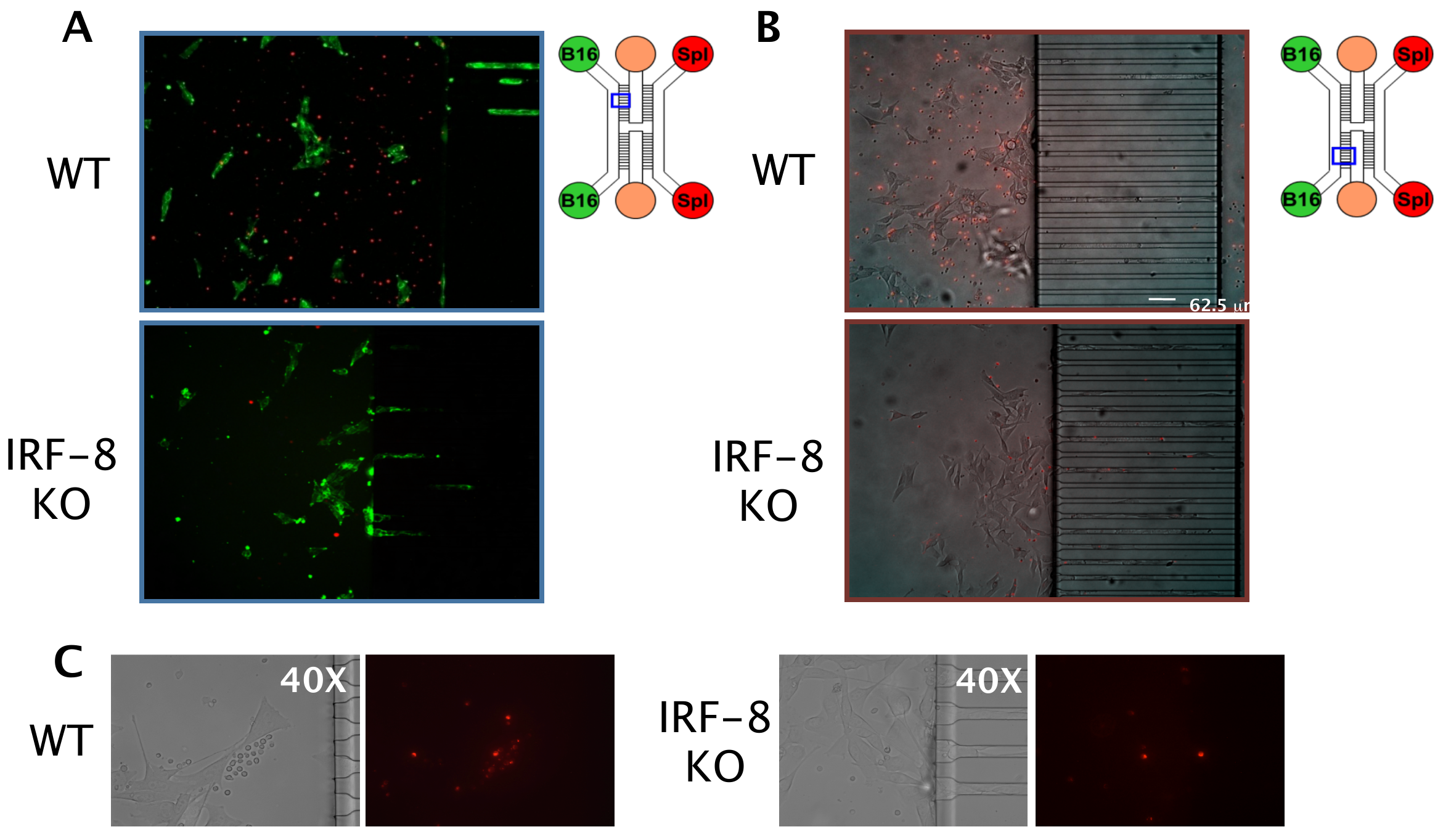}
\caption{Differential migratory behavior of KO and WT cells toward melanoma cells.
B16 melanoma cells (green-labeled) and spleen cells (red-labeled) from either WT or KO mice at 24 h (A) and 48 h (B) after loading onto the microfluidic device.  (C) Detail of splenocytes interacting with B16 cells.}
\label{fig:Real}
\end{figure}

As a result, we collected time series of length $T$ for a number $N$ of distinct migrating lymphocytes, hence for each walk $j, j \in (1,...,N)$ (namely for each single trajectory performed by a white cell) we have the position $(x_i^j,y_i^j)$ at any time step $i, i \in (1,...,T)$ and we can derive the series of step lengths $\{ \Delta x_i \}_{i=1,...,T}$ and $\{ \Delta y_i \}_{i=1,...,T}$.
\begin{figure}[h!]
\includegraphics[width=0.46\textwidth]{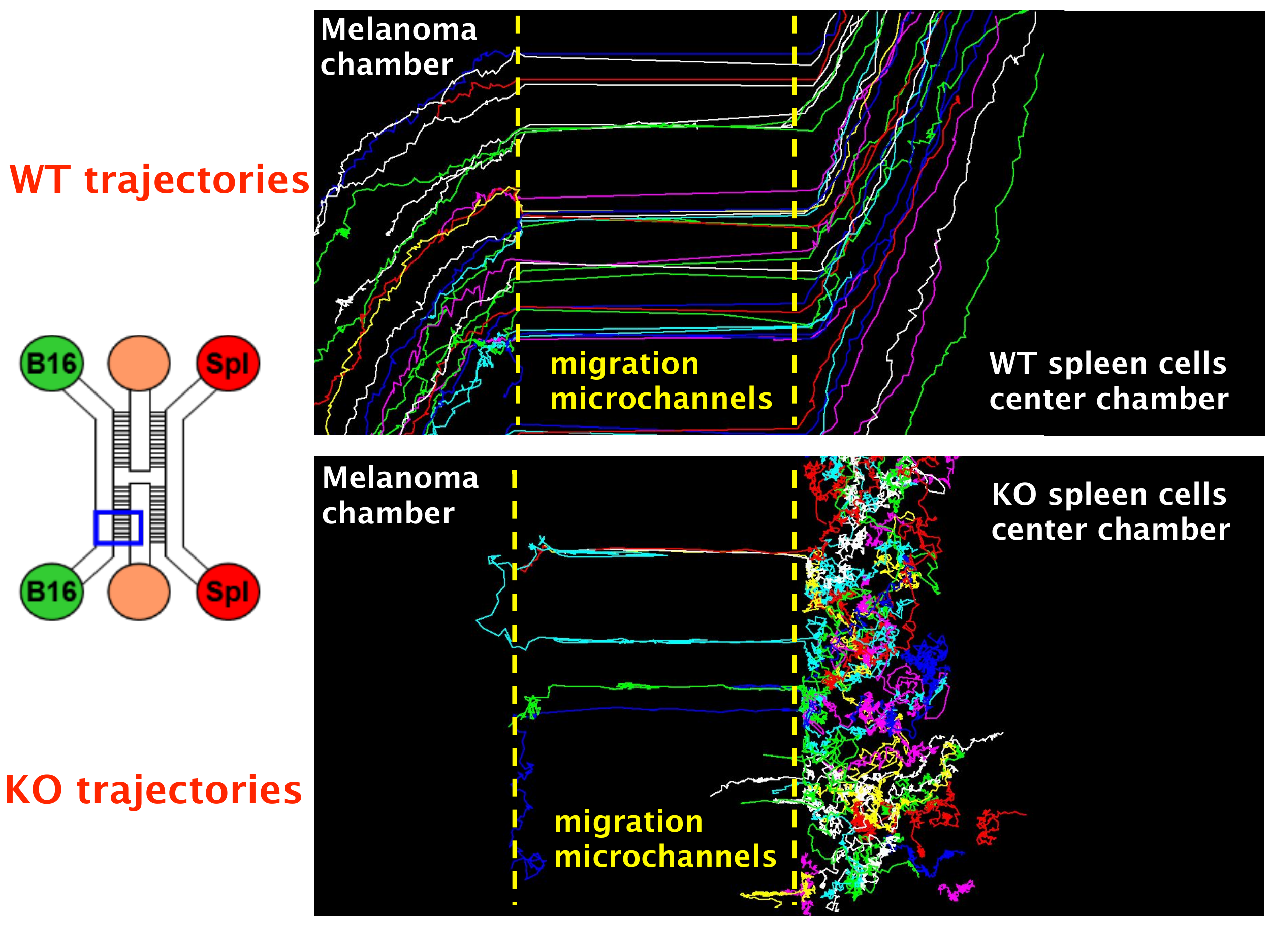}
\caption{Upper panel shows examples of real trajectories performed by WT cells, while lower panel depicts the same for KO cells. WT, but not KO, leukocytes migrate from top-right through bottom-left following a chemokine gradient.}
\label{fig:settings}
\end{figure}

\section{Methods} \label{sec:methods}
In this section we describe the tools used to analyze data on splenocyte exploration and propagation.

As anticipated, data on the positions of splenocytes are updated at a constant rate $\tau = 4 min$ (due to the experimental setting) and measured with precision corresponding to $1$ pixel i.e. $\sim 0.1 \mu m$. Moreover, the splenocytes under investigation perform paths which can, in principle, exhibit some degree of bias (e.g. due to chemical gradients) and some degree of stochasticity (e.g. due to noise).
\newline
Hence, we can model such paths by means of random walks characterized by (synchronized) discrete time steps and moving on a continuous two-dimensional space. This kind of random walk can be described in terms of a probability distribution $p(\mathbf{r},t)$ giving the probability that the walker has covered a distance $\mathbf{r}$ in a time $t$. In fact, we can write
\be
p(\mathbf{r},t + \tau) = \int_{-\infty}^{\infty} p(\mathbf{r'}, t) \psi (\mathbf{r-r'}, t) d\mathbf{r'},
\ee
where $\psi(\mathbf{r-r'}, t)$ is the probability that at time $t$ a step from $\mathbf{r'}$ to $\mathbf{r}$ is performed. For time-homogeneous processes, the width and the direction of a step do not depend on time and the dependence on $t$ can be dropped, i.e. $\psi (\mathbf{r-r'}, t) =  \psi (\mathbf{r-r'})$. Moreover, by exploiting the discreteness of time steps, the time $t$ at which any step occurs is a multiple of $\tau$ in such a way that we can write
\be \label{eq:Master}
p(\mathbf{r},n+1) = \int_{-\infty}^{\infty} p(\mathbf{r'}, n) \psi (\mathbf{r-r'}) d\mathbf{r'},
\ee
$n$ being the number of steps performed up to the time considered and $t= \tau n$.

Of course, the distribution $\psi (\mathbf{r-r'})$ qualitatively controls the resulting random walk, possibly giving rise to deterministic walks (e.g.
$\psi (\mathbf{r}-\mathbf{r'}) = \delta_{\mathbf{r}-\mathbf{r'},\mathbf{k}}$,
$\mathbf{k} \neq \mathbf{0}$,
corresponding to a ballistic motion), to correlated walks (e.g. $\psi(\mathbf{r-r'}) =
f(\mathbf{r \cdot r'})$, where $f$ is a peaked function, corresponding to a motion with a preferred direction), to completely stochastic walks (e.g. $\psi(\mathbf{r}-\mathbf{r'}) = \delta_{|\mathbf{r}-\mathbf{r'}|,\tilde{r}}$, corresponding to an isotropic motion where steps have fixed length $\tilde{r}$), etc. \cite{}.

In Euclidean structures, like the two-dimensional substrate considered here, we can decompose $\mathbf{r}$ into its normal coordinates, i.e. $\mathbf{r} = (x,y)$, and, analogously $\mathbf{r} - \mathbf{r'} = (x-x', y-y') \equiv (\Delta x, \Delta y)$.
Therefore, Eq.~\ref{eq:Master} can be rewritten as
\be\small \label{eq:pp}
p(\mathbf{r},n+1) \equiv p((x,y),n+1) = \int_{-\infty}^{\infty} \int_{-\infty}^{\infty}  p(\mathbf{r'}, n) \psi(\Delta x, \Delta y) dx' dy',
\ee
and, assuming that $\Delta x$ and $\Delta y$ are independent, $\psi(\Delta x, \Delta y)$ can be factorized as $\psi(\Delta x, \Delta y) = \psi_x(\Delta x) \psi_y(\Delta y)$.
%

As suggested by Eq.~\ref{eq:pp}, the knowledge of the specific distribution $\psi(\Delta x, \Delta y)$ possibly allows to get an explicit expression for $p(\mathbf{r},n)$. 

For instance, one can show that \cite{Weiss-1994} any distribution $\psi(\Delta x, \Delta y)$ fulfilling the central limit theorem asymptotically leads to the well-known diffusive limit characterized by the normal distribution
\begin{equation}
p((x,y),t) = \frac{1}{4\pi \mathbf{D} t} e^{-\frac{|\mathbf{r} - \mathbf{v}t|}{4\mathbf{D} t}},
\end{equation}
where $\mathbf{v} = (v_x, v_y)$ accounts for the presence of a drift, while $\mathbf{D}$ is the $2 \times 2$ diffusion-coefficient matrix. In particular, when diffusion is isotropic, i.e. $D_{11}=D_{22}=D$, we have
\be
p((x,y),t) = \frac{1}{4 \pi D t} e^{-\frac{[(x-v_xt)^2 + (y - v_y t)^2]}{4Dt}},
\ee
whose moments are
\begin{eqnarray}
\langle (x,y) \rangle &=& (v_x t , v_y t);\\
\langle x^2 + y^2 \rangle &=& v_x^2 t^2 + v_y^2 t^2 + 4 Dt,
\end{eqnarray}
hence, asymptotically, whenever noise is prevailing, we expect to observe a Brownian motion, i.e. $\mathbf{r} \propto \sqrt{t}$, while, whenever there is a real presence of a drift (signal), we expect ballistic motion, i.e. $\mathbf{r} \propto t$.

Hereafter, we summarize the observables that we are going to analyze, stressing on the kind of information which can be conveyed via their investigation.

\subsection{Step length analysis}

Step lengths are measured in micrometers to quantify the distance covered by the cells during the time-interval among two different (adiacent) frames, i.e. $\tau = 4$ min. The distributions of step lengths immediately provide fundamental information:
\begin{itemize}
\item \emph{Direction dependency.} By means of Pearson's coefficient we can highlight possible correlations between the time series $\{ \Delta x_i\}$ and $\{ \Delta y_i\}$ for a given walk. This analysis also allows to check whether propagations along the two dimensions are independent, namely if $\psi(\Delta x, \Delta y)$ can be (at least approximately) factorized into $\psi_x(\Delta x) \psi_y(\Delta y)$.

\item \emph{Distribution of step lengths.}  If  the distributions display a finite mean $\mu$ and a finite variance $\sigma$, one can apply the central limit theorem to get that $x(n) = \sum_{k=1}^n \Delta x_k$ has average value converging to $n \mu$ with variances scaling like $\sigma \sim\sqrt{n}$, (and analogously along the $y$ direction).
In this case the walk displays a characteristic length scale given by $\mu$.
\end{itemize}

\subsection{Time Correlations}
Time correlations may arise in several different contexts and here we outline two typical cases:
\begin{itemize}
\item \textit{Angular correlation.}
Due to the presence of a forcing field, a diffusive particle may exhibit a preferred direction.
Such a persistence can be measured in terms of the angle $\theta$ between two consequent steps; in particular, to highlight the existence of a short-term memory we consider the temporal angular correlations $C(t)$ defined as
\begin{equation}
C(t)=\left \langle \overline{\cos[\theta(t^{\prime}+t)-\theta(t^{\prime})] } \right \rangle,
\label{corrang}
\end{equation}
%
%
where the average $\bar{\cdot}$ is performed over $t^{\prime}$ and the average $\langle \cdot \rangle$ is performed over the set of random walks. Hence,  $C \sim 0$ implies isotropy, which, in turn, implies that migrating white cells are not pointing to any specific target; conversely, $C \neq 0$ is a necessary requisite in order to keep a coordinate motion toward the target (melanoma cells in the present case).

\item \textit{Acceleration phenomena}.
In order to figure out the possible existence of slowing down and/or speeding up phenomena one can consider the mean step length $l_k= \sqrt{\Delta x_k^2 + \Delta y_k^2}$ of a single random walk at each time step $k$ and calculate
\begin{equation}
\bar{l}(n)=\sum_{k=1}^{n} \frac{l_k}{k}
\label{passomedio}
\end{equation}
Hence, at each time, one obtains the average of the steps taken up to the $n$-th time step by the single cell.
A (at least approximately) constant behavior of $\bar{l}(n)$ ensures that, independently of the instant of time and of the place where the cell is currently located, the length of the step taken tends to remain the same. On the contrary, in the case of an acceleration/deceleration, an increasing/decreasing behavior of $\bar{l}(n)$ is expected.
\end{itemize}

%
%
%
%
%
%
%

The quantities described so far focus on ``microscopic'' features of a walk as they imply a fine zoom on the walk itself. From such a description one is usually able to derive the ``macroscopic'' behavior, typically measured in terms of the average distance $\langle r (t) \rangle$ covered as a function of time, whose following observables are due to.

\subsection{Mean displacement}

One can consider both the displacements $x(t)$ and $y(t)$ along the $x$ and $y$ axes and the overall distance
\begin{equation}
r(t)=\sqrt{[x(t)]^2+[y(t)]^2}.
\label{pitagora}
\end{equation}
Notice that $x(t)$ and $y(t)$ are calculated with respect to the initial point in order to get the effective displacement\footnote{The rigorous notation for $r(t)$ should be $r(t)=\sqrt{[x(t) - x(0)]^2+[y(t) - y(0)]^2}$ in order to account for the effective displacement with respect to the original position. However, the notation has been lightened throughout the paper.}.
All these quantities can be computed for every cell at each time and, then, they are averaged over the ensemble of cells, obtaining $\left \langle x(t) \right \rangle$, $\left \langle y(t) \right \rangle$ and $\left \langle r(t) \right \rangle$.
In fact, the latter quantities represent the mean displacement of the system as a whole.

Now, the scaling of $\left \langle r(t) \right \rangle$ with respect to $t$ is often used to qualitatively define the kind of diffusion. For instance $\left \langle r(t) \right \rangle \sim \sqrt{t}$ is typical of simple diffusion, a linear law $\left \langle r(t) \right \rangle \sim t$ is typical of drifted motion, while a power law $\left \langle r(t) \right \rangle \sim t^{\alpha}$ is referred to as anomalous diffusion emerging, for instance, in the presence of crowded environment and/or fractal substrates \cite{}.


\subsection{Tortuosity}

When dealing with the movement of a biological particle one is often interested in the tortuosity of its path, namely in how twisted the path is in a given space or time \cite{alm}. Clearly, this is related to the mean displacement: highly tortuous paths will spread out in space slowly, while straight paths will spread out in space quickly. Hence, it can be useful to measure and study the tortuosity of observed paths in order to understand the processes involved, estimate the area spanned by a cell and predict spatial dispersal.

Tortuosity can be quantified by comparing the overall net displacement of a path with the total path length. For example, if a random walk starts at location $(0,0)$ and, after $n$ steps with lengths $l_j$ $(j=1,\dots,n)$, ends at $(x_n,y_n)$, then we can measure the so-called straightness index $S$ as \cite{Batschelet-1981}
\begin{equation}
S(n)= \frac{\sqrt{[x(n)]^2+ [y(n)]^2}}{\sum_{j=1}^{n} \Delta r(j)},
\label{straightness}
\end{equation}
which ranges in $0$ and $1$, where $1$ corresponds to movement in a straight line (the shortest distance between two points in the two dimensional Euclidean space the LabOnChip has built on) and $0$ corresponds to a returning (thus tortuous) path.

\subsection{Ergodicity}

In the context of stochastic processes, we define ergodic a system where time and ensemble averages converge \cite{}. Simple Brownian motion owns this property. Hence, the possible non ergodicity of the system is a measure of how large is the deviation of the process examined from a normal diffusion. For example, when cells diffuse, one can find that the time averages vary from one cell to the next \cite{bar}.
A convenient way to check ergodicity is the comparison between the mean square displacement (MSD) of diffusing particles and the time-averaged MSD defined as:

\begin{equation}
\overline{\delta^2 (t)} = \frac{1}{T-t} \int_{0}^{T-t} [r(t'+t)-r(t')]^2 dt'.
\label{delta}
\end{equation}
%
\newline
In the case of Brownian motion in two dimensions, we have that
\begin{equation}
\lim_{t \to \infty} \overline{\delta^2} = 4D t
\end{equation}
is precisely the same as the MSD averaged over a large ensemble of particles,
\begin{equation}
\left \langle r^2 \right \rangle = 4D t,
\end{equation}
which quantitatively confirms that in the Brownian motion the ergodicity is preserved. In particular, in this process, a measurement of $\overline{\delta^2}$ and, therefore, $D$ in the time interval ($0,t$) will be identical to a measurement in the interval ($t,2t$) for large $t$. Therefore, if a system shows the ergodicity property, it surely respects the time-translational invariance, which, instead, is not applicable in many kinds of anomalous diffusion, such as subdiffusion processes \cite{bar}. \\

\section{Results}
In this section we describe the results obtained with our analysis of the KO and WT splenocytes.

\subsection{Knock-out splenocytes: a simple random walk}

\begin{figure}
\centering
\includegraphics[width=8cm]{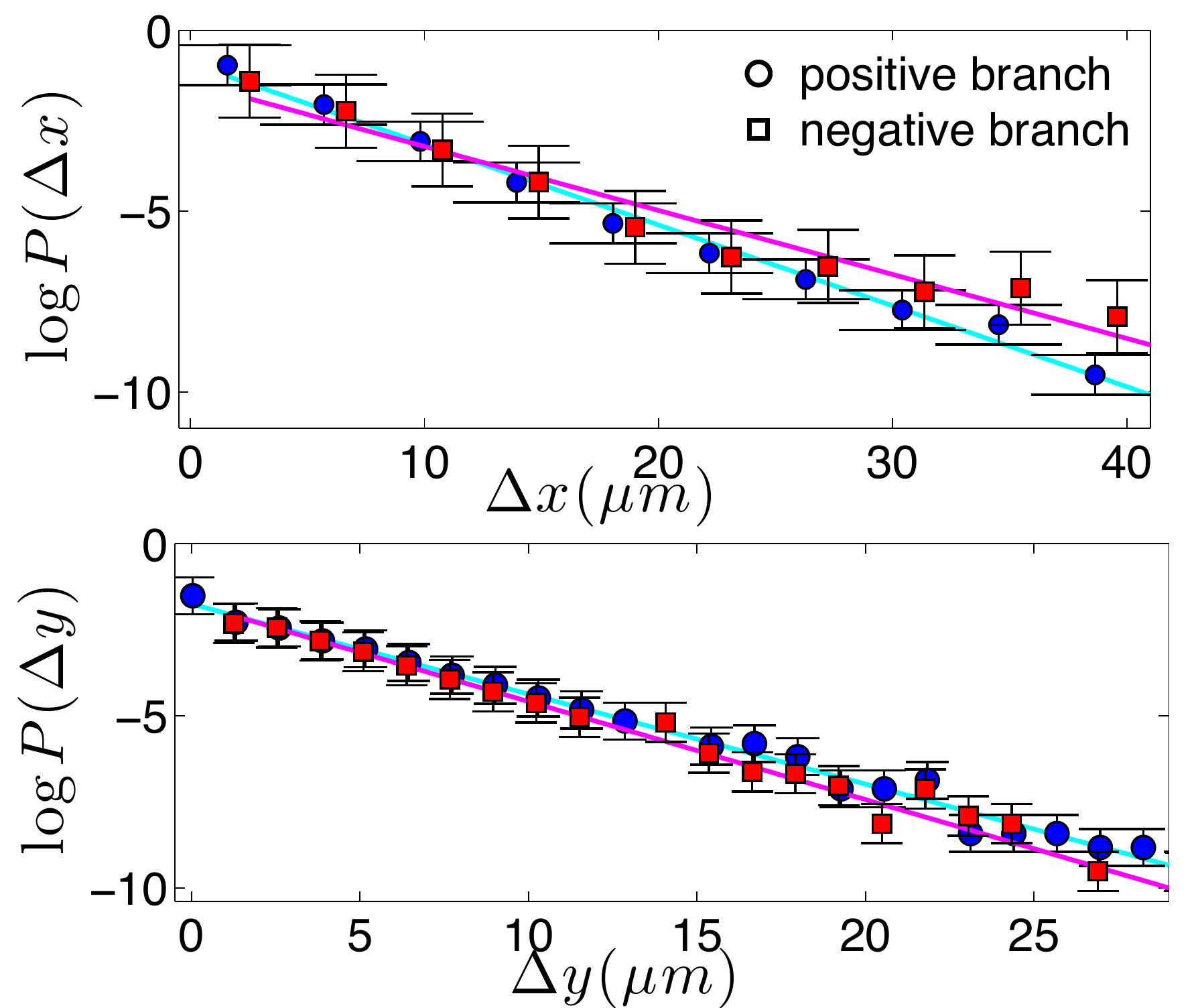}
\caption{Logarithm of the probability distribution for the step length along the $x$ direction (upper panel) and along the $y$ direction (lower panel). Experimental data (symbols) with standard errors are fitted by the exponential distribution (solid line) given by Eq.~\ref{eq:expo}. All fits display $R^2 \approx 0.99$. The best fit coefficients are reported in Tab.~\ref{step_length}, where a comparison between average values from experiments and theoretical description is also provided. }
\label{DeltaKO}
\end{figure}

Since KO splenocytes were poorly reactive to melanoma cells, almost no  cell was able to get into the micro-channels. Thus, the motion of these cells was studied only in the center chamber. \\
The available data were filtered with the compromise of obtaining the positions of splenocytes monitored from the same instant of time $t_0$ (and not to include splenocytes initially too close to the channel wall, in order to avoid collisions with it, which could distort results) and, at the same time, of getting a reasonable statistics with the minimum number of analyzed cells.
From this selection procedure we outlined 30 splenocytes for our analysis: It is remarkable that with such a small number of elements the statistics were already very significant (as shown below). Implication on collective capabilities of leukocytes will be discussed at the end.

As anticipated, our analysis begins with the determination of microscopic quantities.

First, we notice that there is no manifest spatial correlation between $\Delta x$ and $\Delta y$ along a single walk. Also, the histogram of the Pearson coefficient $\rho_{\Delta x \Delta y}$ for each walk peaks at zero (not shown).
As for the distribution of step lengths $\psi_x(\Delta x)$ and $\psi_y(\Delta y)$, we find that, at each time step, splenocytes perform a jump whose width is stochastic and exponentially distributed, as shown in Fig.~\ref{DeltaKO}. In particular, for both directions (displacements along $x$ and along $y$) and for both branches (negative and positive displacements)
the best fit is given by
\be \label{eq:expo}
P(x) = \lambda e^{- \lambda x};
\ee
best fit coefficients, summarized in Tab.~\ref{step_length}, are consistent with the experimental average values and highlight overall, within the error, a good symmetry. This suggests that KO cells are not pointing to any target.

Moreover, the exponential distribution clearly satisfies the central limit theorem and this rules out the existence of L\'evy flights among KO splenocytes. In other words, these splenocytes proceed smoothly and with rather regular steps.


\begin{table}[!htbp]\small
\begin{center}
\begin{tabular}{c | c | c | c | c}
\hline
\hline
$Branch$ & $\lambda_{x}^{-1} \, [\mu m]$ &  $\left \langle \Delta x \right \rangle \, [\mu m]$ & $\lambda_{y}^{-1} \, [\mu m]$ & $\left \langle \Delta y \right \rangle \, [\mu m]$ \\
\hline
Positive & 4.5  $\pm$ 0.1 &  4.5  $\pm$ 0.8 & 4.3  $\pm$ 0.1 &  4.3  $\pm$ 0.9 \\
Negative & -4.5 $\pm$ 0.1 &  -4.5 $\pm$ 0.9 & -4.3 $\pm$ 0.1 &  -4.3 $\pm$ 0.7 \\
\hline
\end{tabular}
\quad
\caption{Characteristic step length for KO splenocytes along the $x$ and $y$ axes. For each axis we compare the mean value of the exponential fit $\lambda^{-1}$ with the average over all values for $\Delta x$ and $\Delta y$.}
\label{step_length}
\end{center}
\end{table}

\begin{figure} [!htbp]
\includegraphics[width=7.5cm]{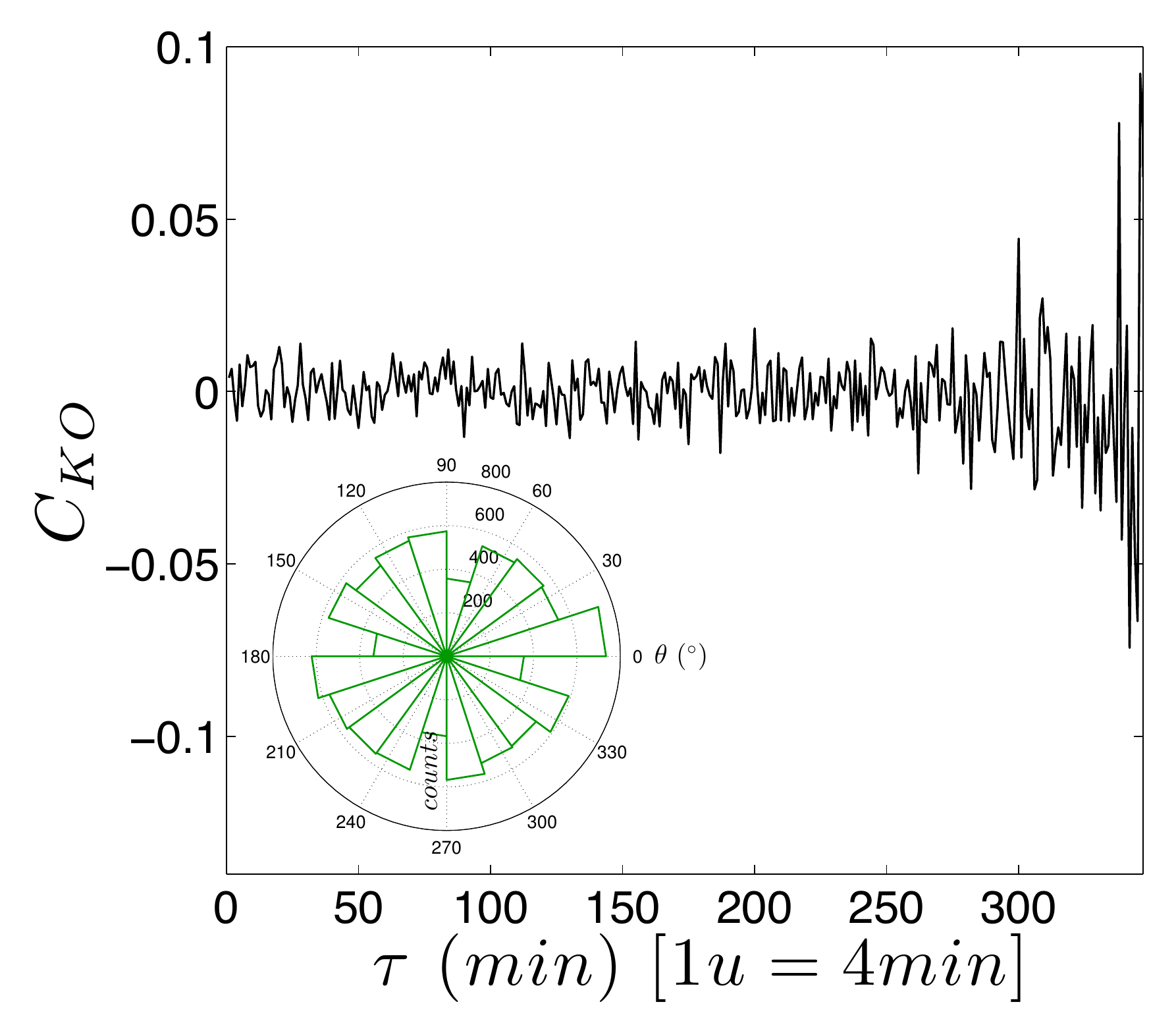}
\caption{Inset: polar histogram of the turning angle. The distribution has zero mean, hence, no angular correlation is observed.
Main plot: angular correlation function $C_{\textrm{KO}}(t)$ of the turning angle $\theta$. This correlation function shows fast oscillations resulting from continuous change of direction by each KO cell.}
\label{corr_ang_KO}
\end{figure}

Let us now consider the turning angle $\theta$ between two consecutive steps.
The distribution of $\theta$ over the whole set of walks and the related time correlation $C(t)$, (see Eq.~\ref{corr_ang_KO}) are shown in Fig.~\ref{corr_ang_KO}: The turn amplitude has zero mean, implying again that every choice of direction is not correlated with the previous one and the motion is isotropic.
Moreover, $C(t)$ has zero average, confirming that there is no connection between the direction of a step with that of the following one so that one can exclude the presence of memory  or collective organization in the process.

\begin{figure} [!htbp]
\centering
\includegraphics[width=8cm]{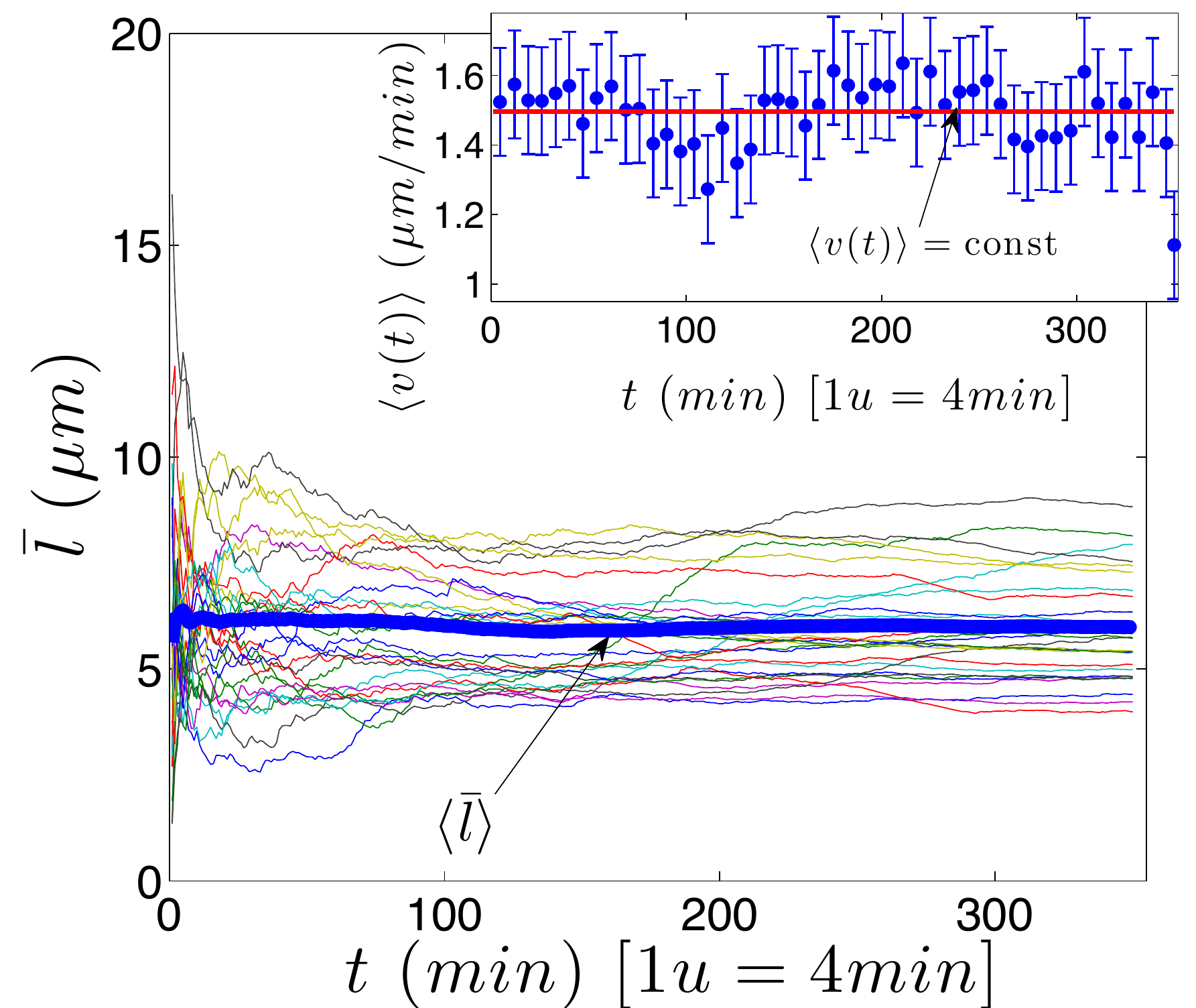}
\caption{Inset: binned data ($\bullet$) with standard errors of mean instantaneous speed and related best fit (solid line). KO splenocytes do not slow down over the observation time of 350 frames. Thus, for practical purposes, we can consider them as being in a time-independent state. Main plot: mean step length $\bar{l}$ for each KO cell (thin curve) and mean step length averaged over all splenocytes (thick curve) at each time, which is essentially constant.}
\label{v_KO}\label{corr_temp_KO}
\end{figure}

Finally, we do not find any significant temporal correlation among steps since, for each cell, the step $\overline{l}$ (see equation $\ref{corr_temp_KO}$) converges a constant value; as shown in Fig.$\ref{corr_temp_KO}$, no acceleration is observed in the process and the instantaneous speed $\langle v(t) \rangle$ is stable. More precisely, it fluctuates around $1.5 \pm 0.1$ $\mu m/min$ in agreement with the results of other \textit{in vitro} experiments, showing that, in the absence of external gradient guiding the KO splenocytes, these move with an average speed $1 \sim 4$ $\mu m/min$ \cite{yang}.

%

Thus, from this microscopic analysis we can confidently derive that KO splenocytes move rather uniformly and isotropically, with no manifest
persistence or bias, consistently with the expected lack of collective organization.

\begin{figure} [htbp]
\centering
\includegraphics[width=8.0cm]{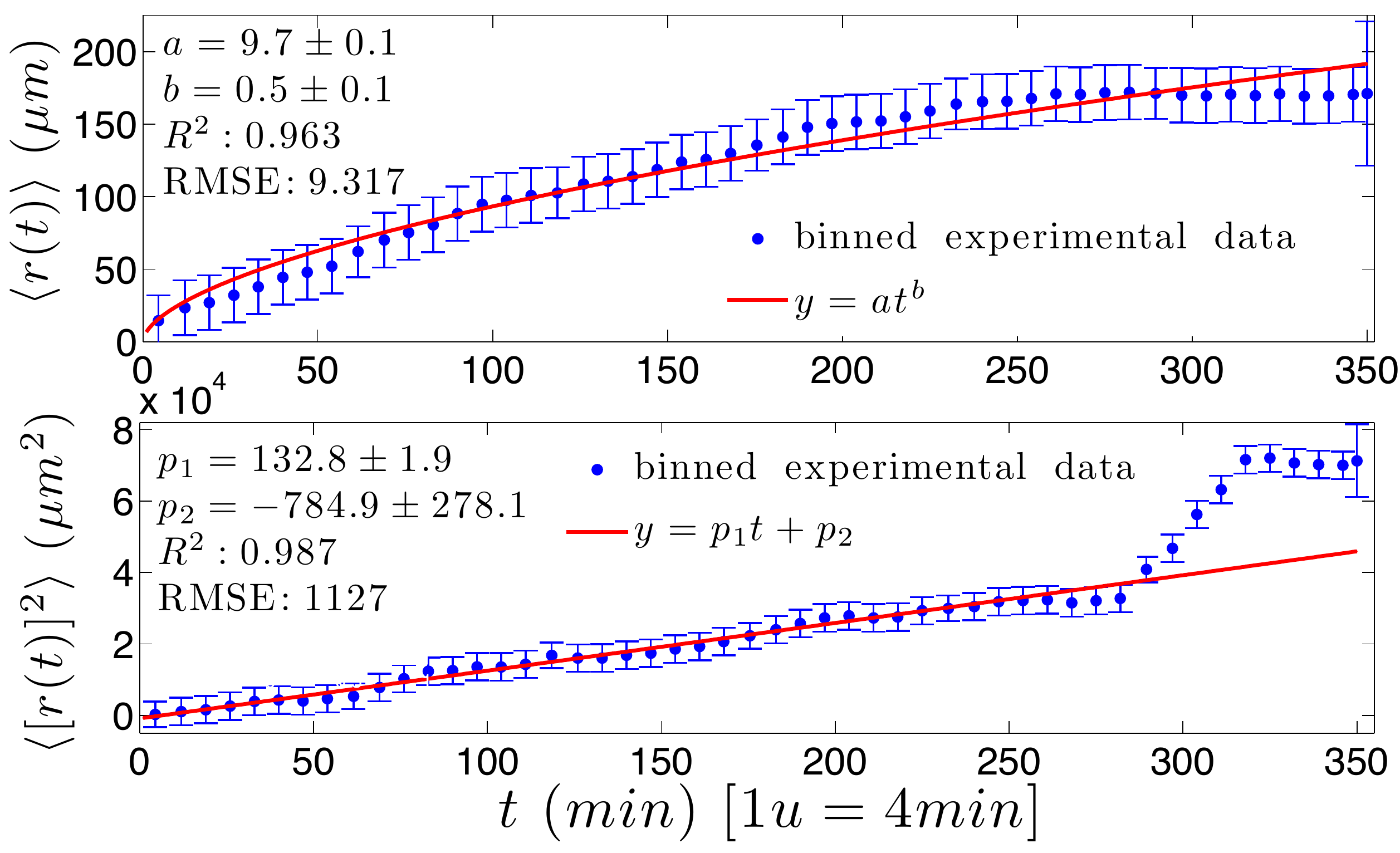}
\caption{Main displacement $\left \langle r(t) \right \rangle$ (upper panel) and mean squared displacement $\left \langle [r(t)]^2 \right \rangle$ (lower panel) for KO splenocytes. Experimental (binned) data ($\bullet$) with a standard errors are compared with best fits (solid line) whose coefficients are properly shown. At late times the mean walk displays deviations with respect to the typical behavior: this effect is due to the reflection (or partial collision)  of cells at channels; since just a fraction of cells is expected to sense the wall, the net effect is that the distribution of displacements becomes broader and the mean squared displacement is increased, while the mean displacement slows down.
}
\label{r_KO}
\end{figure}

As for the macroscopic analysis, we show the time evolution of the mean distance $\left \langle r(t) \right \rangle$ and of the mean-square displacement $\left \langle r^2(t) \right \rangle$ (see Fig.~\ref{r_KO}), which are proportional to $\sqrt{t}$ and to $t$, respectively.
This is the typical behavior of pure diffusion (see Eq.~\ref{pitagora}), in agreement with the results above.


\begin{figure} [!htbp]
\centering
\includegraphics[width=8cm]{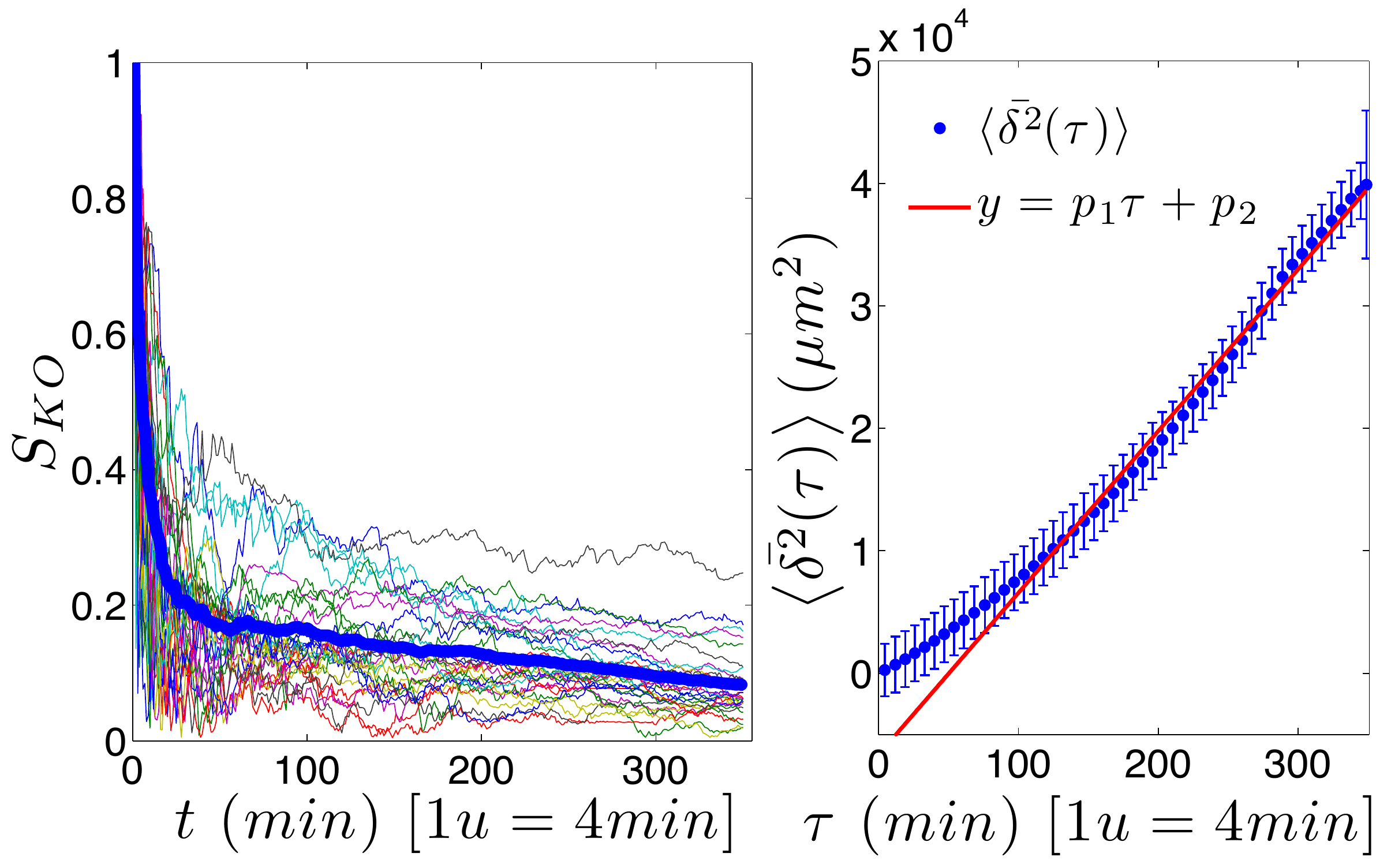}
\caption{Left panel: straightness index $S_{KO}$ for each KO lymphocyte (thin curves) and straightness index averaged over all splenocytes (thick curve) at each time.
Right panel: binned data for $\langle \overline{\delta^2 (\tau)} \rangle$ ($\bullet$) with standard errors for KO splenocytes and best linear fit (solid line). The angular coefficient of $\langle \overline{\delta^2 (\tau)} \rangle$ is $p_1= 1.20 \pm 0.01$.
}
\label{S}
\label{tuttidelta_KO}
\end{figure}

From the whole set of results described so far we can consistently derive that KO cells perform a simple random motion (at least as for their free path) and that any anomalous diffusion can be excluded in this context.
This can be further corroborated by the straightness index $S$ (see equation $\ref{straightness}$), which, as shown in Fig.~\ref{S}, decreases rapidly over time, approaching to zero (see section $\ref{sec:methods}$). Indeed, in a normal diffusion process, the tortuosity of the path is high, because the particles do not move in a specific direction, but tend to explore the space rather compactly. For this reason, a simple Brownian motion spreads more slowly than a random walk with bias (as described in the next section).

Finally we consider the ergodicity problem. We measure for each trajectory the time average $\overline{\delta^2 (\tau)}$ (see equation $\ref{delta}$), which is then averaged over all trajectories to get $\langle \overline{\delta^2 (\tau)} \rangle$. As shown in Fig.~\ref{tuttidelta_KO} the angular coefficient of $\langle \overline{\delta^2 (\tau)} \rangle$ is approximately $1$, which is typical of a Brownian motion, since $\overline{\delta^2 (\tau)} \sim t$. 

More precisely, $\langle \overline{\delta^2 (\tau)} \rangle$ and $\left \langle r^2(t) \right \rangle$ have the same linear shape with comparable slope within the error (see Tab.~\ref{ang_coeff}).



\begin{table}[!htbp]
\begin{center}
\begin{tabular}{c | c }
\hline
\hline
Quantities & Angular coefficient \\
\hline
$\left \langle \overline{\delta^2 (t)} \right \rangle$ & 132.0 $\pm$ 1.3 \\
$\left \langle r^2(t) \right \rangle$ &  132.8 $\pm$ 1.9 \\
\hline
\end{tabular}
\quad
\caption{Angular coefficients for $\langle \overline{\delta^2 (t)} \rangle$ and $\left \langle [r(t)]^2 \right \rangle$.}
\label{ang_coeff}
\end{center}
\end{table}
Thus, $\langle \overline{\delta^2 (t)} \rangle \rightarrow 4D t$ and there is equivalence of time and ensemble average, which is the hallmark of ergodicity.
In particular, both procedures agree on the estimate of the diffusion coefficient, which turns out to be approximately $D \sim 8 \mu m^2/min$.

In conclusion, the behavior of KO splenocytes can be characterized  by a simple random walk, hence with a manifest lack of collective organization. In this respect, this result confirms the important role played by IRF-8 as a central regulator of immune response and anticancer immunosurveillance \cite{neoplasia}.

\subsection{Wild type splenocytes: a biased random walk}

Since WT splenocytes express IRF-8, they are expected to have a competent response to the tumor: Indeed, as we will see, WT  splenocytes migrate towards B16 melanoma cells, in the attempt to contain their expansion.

Point-by-point tracking, between $24$ and $48$ hours from the beginning of the experiment, showed that WT splenocytes are endowed with the potential ability to cross the microchannels connecting the central channel with the melanoma channel (left channel). Thereafter, the performances of WT splenocytes in the right channel, i.e., before passing the microchannels (WT-PRE), and of WT splenocytes in the left channel, i.e., after passing the microchannels (WT-POST), are treated separately.



\begin{figure}
\includegraphics[width=8cm]{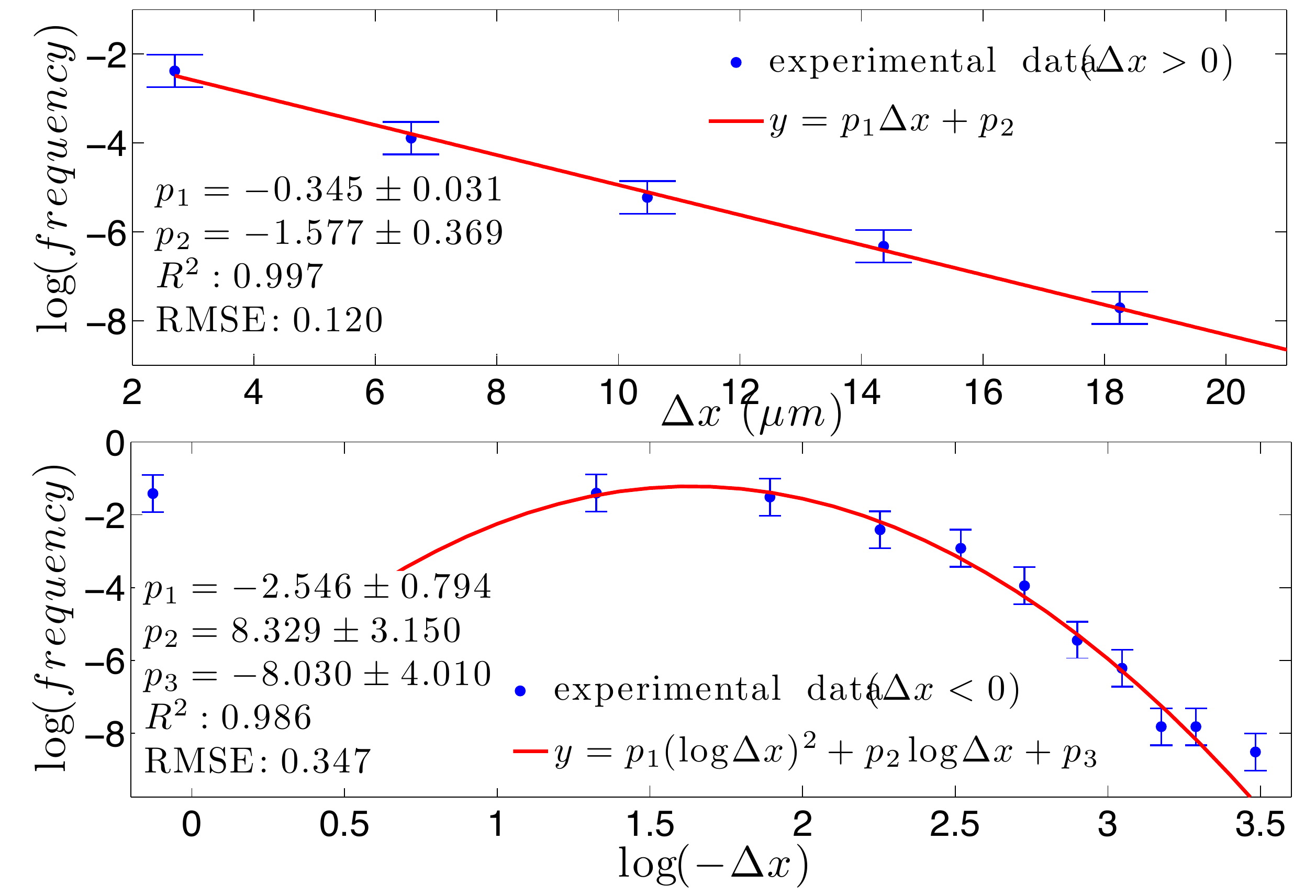}
\caption{$P(\Delta x)$ of WT-PRE splenocytes along the positive direction (upper panel, notice the semilogarithmic scale) and along the negative direction (lower panel, notice the logarithmic scale). Analogous plots are obtained for the $y$ direction (not shown). Experimental data of the distributions ($\bullet$) with standard errors are compared with best fits (solid line). Note that the distribution is broadened along the negative $x$ direction (and, analogously, along the positive $y$ direction).}
\label{Delta_WTPRE}
\end{figure}

\begin{figure}
\includegraphics[width=8cm]{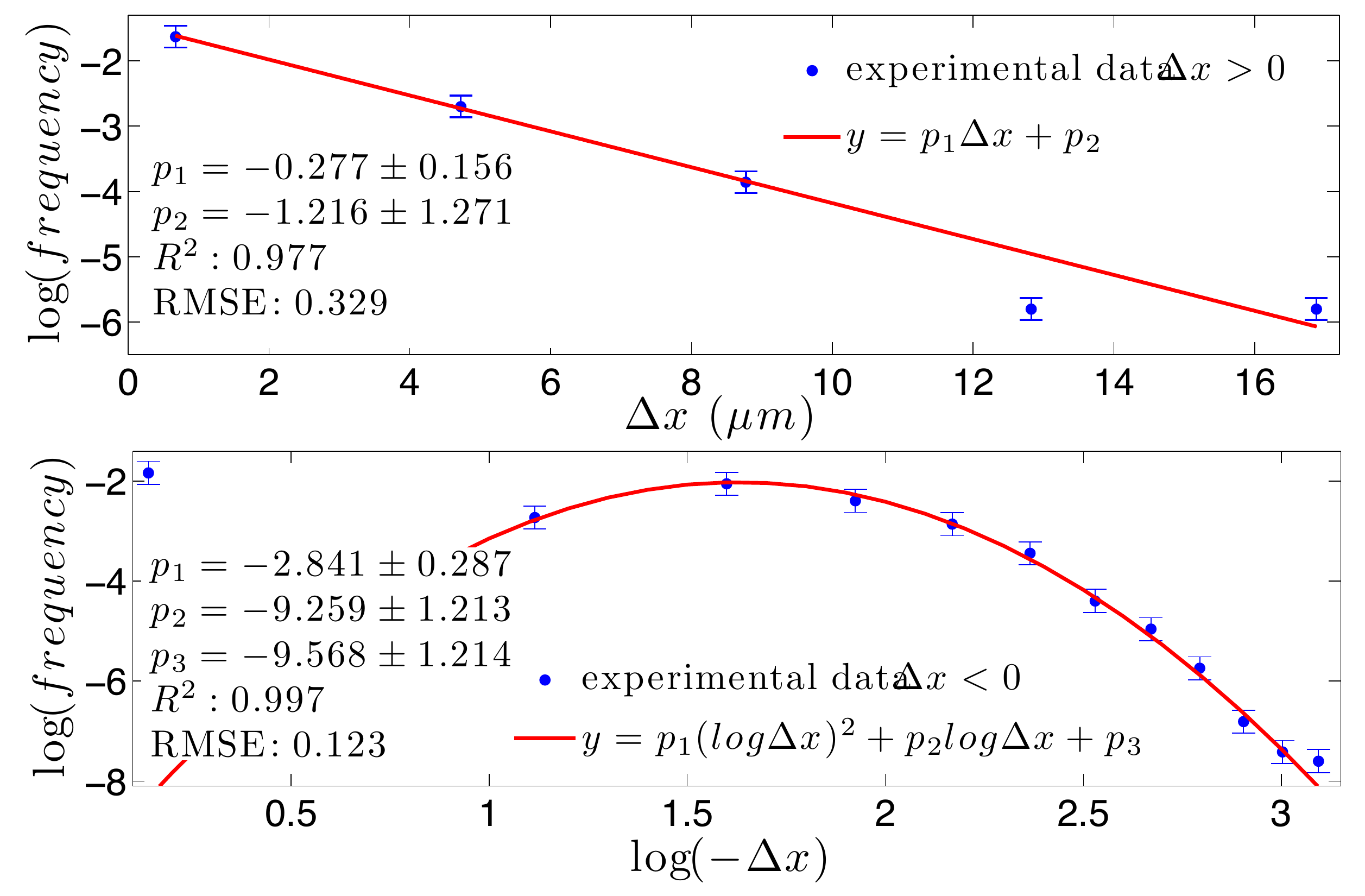}
\caption{$P(\Delta x)$ of WT-POST splenocytes along the positive direction (upper panel, notice the semilogarithmic scale) and along the negative direction (lower panel, notice the logarithmic scale). Analogous plots are obtained for the $y$ direction (not shown). Experimental data of the distributions ($\bullet$) with standard errors are compared with best fits (solid line). Note that the distribution is broadened along the negative $x$ direction (and, analogously, along the positive $y$ direction).}
\label{Delta_WTPOST}
\end{figure}

First, we consider the probability distributions of step lengths $P(\Delta x)$ and $P(\Delta y)$. Interestingly, a qualitative difference with respect to the case of KO splenocytes emerges: along the direction pointing to melanoma cells (i.e. along negative $x$ and positive $y$ directions) distributions are broadened and best-fits are now provided by log-normal distributions (see Fig.$\ref{Delta_WTPRE}$ and Fig.$\ref{Delta_WTPOST}$, lower panels); on the other hand along the opposite direction (i.e. along positive $x$ and negative $y$ directions)
distributions are still exponential (see Fig.$\ref{Delta_WTPRE}$ and Fig.$\ref{Delta_WTPOST}$, upper panels).
This constitutes a clear evidence of the ability of perceiving the presence of a chemotactic gradient along negative $x$ and positive $y$ directions which, on the cartesian $xy$ plane, corresponds to a drift towards the second quadrant, where the source of melanoma cells resides. 
This kind of behavior is evidenced for both WT-PRE and WT-POST; related fitting coefficient are reported in 
Tabs.~$3$-$6$: notice that for WT-PRE the effect is stronger.
This may be due to the fact that WT-POST splenocytes, being in the left channel, are at least partially surrounded by tumor cells in such a way that the resulting signaling is less focussed and, consequently, this drift becomes weaker.

\begin{table}[!htbp] \label{tab:1}\small
\begin{center}
\begin{tabular}{c | c | c }
\hline
\hline
Positive branch & $\lambda^{-1}_{x} \, [\mu m]$ & $\left \langle \Delta x \right \rangle \, [\mu m]$ \\
\hline
$\Delta x$ WT-PRE & 2.9 $\pm$ 0.3 &  3.4  $\pm$ 0.4 \\
$\Delta x$ WT-POST & 3.6 $\pm$ 2.0 & 4.1 $\pm$ 0.2 \\
\hline
\end{tabular}
\caption{Characteristic step length for WT-PRE and WT-POST splenocytes along the positive $x$ direction. For both groups of splenocytes we compare the mean value of the exponential fit $p_{1x}^{-1}$ with the average over all values for positive $\Delta x$.}
\label{step_length_WT_XPos}
\end{center}
\end{table}

\begin{table}[!htbp]\label{tab:2}\small
\begin{center}
\begin{tabular}{c | c | c | c | c}
\hline
\hline
Neg. branch & $\mu_{x} \, [\mu m]$ & $\left \langle \Delta x \right \rangle \, [\mu m]$ & $\sigma_{x} \, [\mu m]$ & $\sigma_{x, exp} \, [\mu m]$\\
\hline
$\Delta x$ WT-PRE & 6.2 $\pm$ 1.9 &  8.0 $\pm$ 0.5 & 3.2 $\pm$ 0.4 & 5.0 $\pm$ 0.5\\
$\Delta x$ WT-POST & 6.1 $\pm$ 1.8 & 6.5 $\pm$ 0.2 & 3.0 $\pm$ 0.7 & 4.2 $\pm$ 0.2\\
\hline
\end{tabular}
\caption{Characteristic step length for WT-PRE and WT-POST splenocytes along the negative $x$ direction. For both groups of splenocytes we compare the mean value $\mu_x$ of negative $\Delta x$ obtained by the log-normal fit with the experimental average and the standard deviation $\sigma_{x}$ with the experimental value.}
\label{step_length_WT_XNeg}
\end{center}
\end{table}
\begin{table}[!htbp] \label{tab:3}\small
\begin{center}
\begin{tabular}{c | c | c | c | c}
\hline
\hline
Pos. branch & $\mu_{y} \, [\mu m]$ & $\left \langle \Delta y \right \rangle \, [\mu m]$ & $\sigma_{y} \, [\mu m]$ & $\sigma_{y,exp} \, [\mu m]$\\
\hline
$\Delta y$ WT-PRE & 10.1 $\pm$ 2.2 &  12.7 $\pm$ 0.5 & 3.9 $\pm$ 0.4 & 5.9 $\pm$ 0.5\\
$\Delta y$ WT-POST & 6.2 $\pm$ 2.0 & 4.0 $\pm$ 0.2 & 3.4 $\pm$ 0.3 & 5.8 $\pm$ 0.2 \\
\hline
\end{tabular}
\caption{Characteristic step length for WT-PRE and WT-POST splenocytes along the positive $y$ direction. For both groups of splenocytes we compare the mean value $\mu_y$ of positive $\Delta y$ obtained by the log-normal fit with the experimental average and the standard deviation $\sigma_{y}$ with the experimental value.}
\label{step_length_WT_YPos}
\end{center}
\end{table}
\begin{table}[!htbp] \label{tab:4}\small
\begin{center}
\begin{tabular}{c | c | c }
\hline
\hline
Negative branch & $\lambda^{-1}_{y} \, [\mu m]$ & $\left \langle \Delta y \right \rangle \, [\mu m]$ \\
\hline
$\Delta y$ WT-PRE & 2.7 $\pm$ 1.7 &  2.7  $\pm$ 1.0 \\
$\Delta y$ WT-POST & 3.7 $\pm$ 0.7 & 3.6 $\pm$ 0.3 \\
\hline
\end{tabular}
\quad
\caption{Characteristic step length for WT-PRE and WT-POST splenocytes along the negative $y$ direction. For both groups of cells we compare the mean value of the exponential fit $p^{-1}_{1y}$ with the average over all values for negative $\Delta y$.}
\label{step_length_WT_YNeg}
\end{center}
\end{table}

\begin{figure} [htbp]
\centering
\includegraphics[width=9cm]{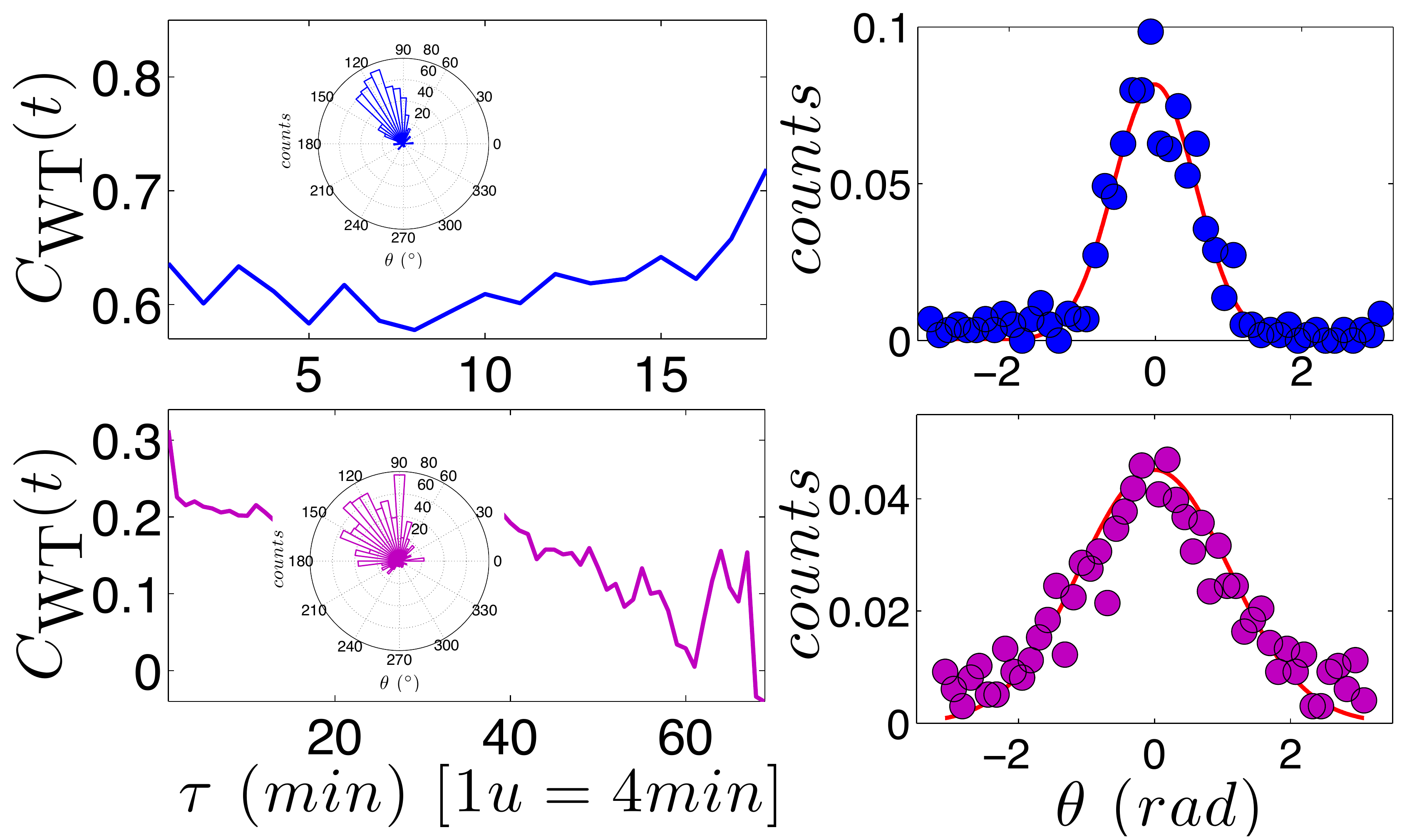}
\caption{Left panels: angular correlation function $C_{\textrm{WT}}(t)$ of the turning angle $\theta$ of WT-PRE splenocytes (upper panel) and of WT-POST splenocytes (lower panel). Each turn occours every 4 minutes. In both cases the correlation is mostly positive, meaning that the process has memory. In the inset we show the polar histrogram of the angle measured with respect to the horizontal axis. Right panels: distribution of the turning angle for WT-PRE splenocytes (upper panel) and of WT-POST splenocytes (lower panel). The experimental distribution is fitted by a Gaussian peaked at $0$ $rad$, since splenocytes tend to maintain the same direction.}
\label{C_tau_WT}
\label{hist_ang_WT}
\end{figure}

The data described so far suggest that WT splenocytes can be modeled by biased random walks. This is indeed corroborated by the distribution of the turning angle $\theta$ and by the related angular correlation $C_{\mathrm{WT}}(t)$ (see Fig.~\ref{hist_ang_WT}).
The bias is especially strong for WT-PRE. Indeed, as mentioned above, in the left channel, splenocytes tend to change direction slighlty more frequently because of a broadened presence of melanoma cells.


\begin{figure} [!htbp]
\includegraphics[width=6.5cm]{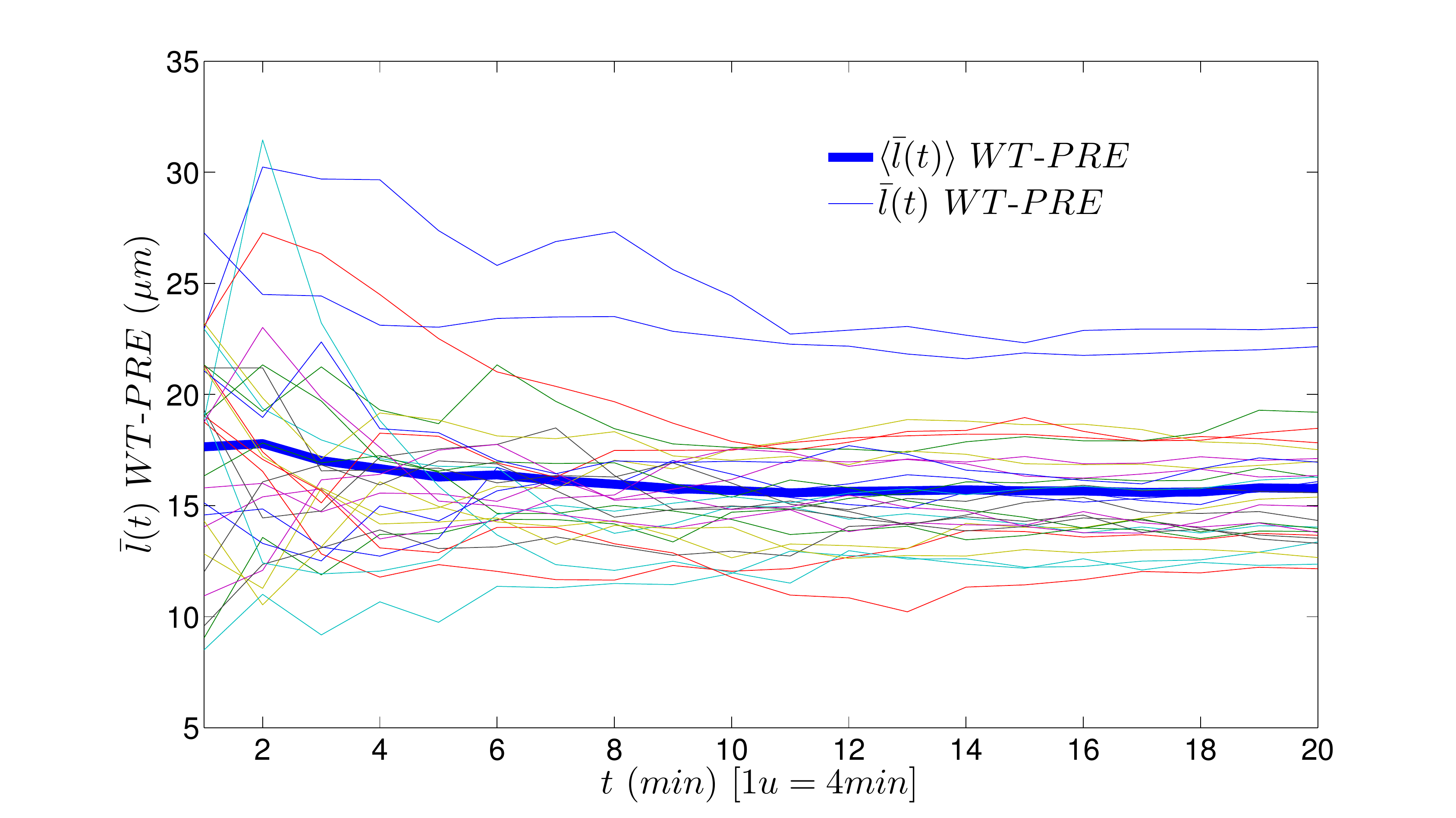}
\includegraphics[width=6.5cm]{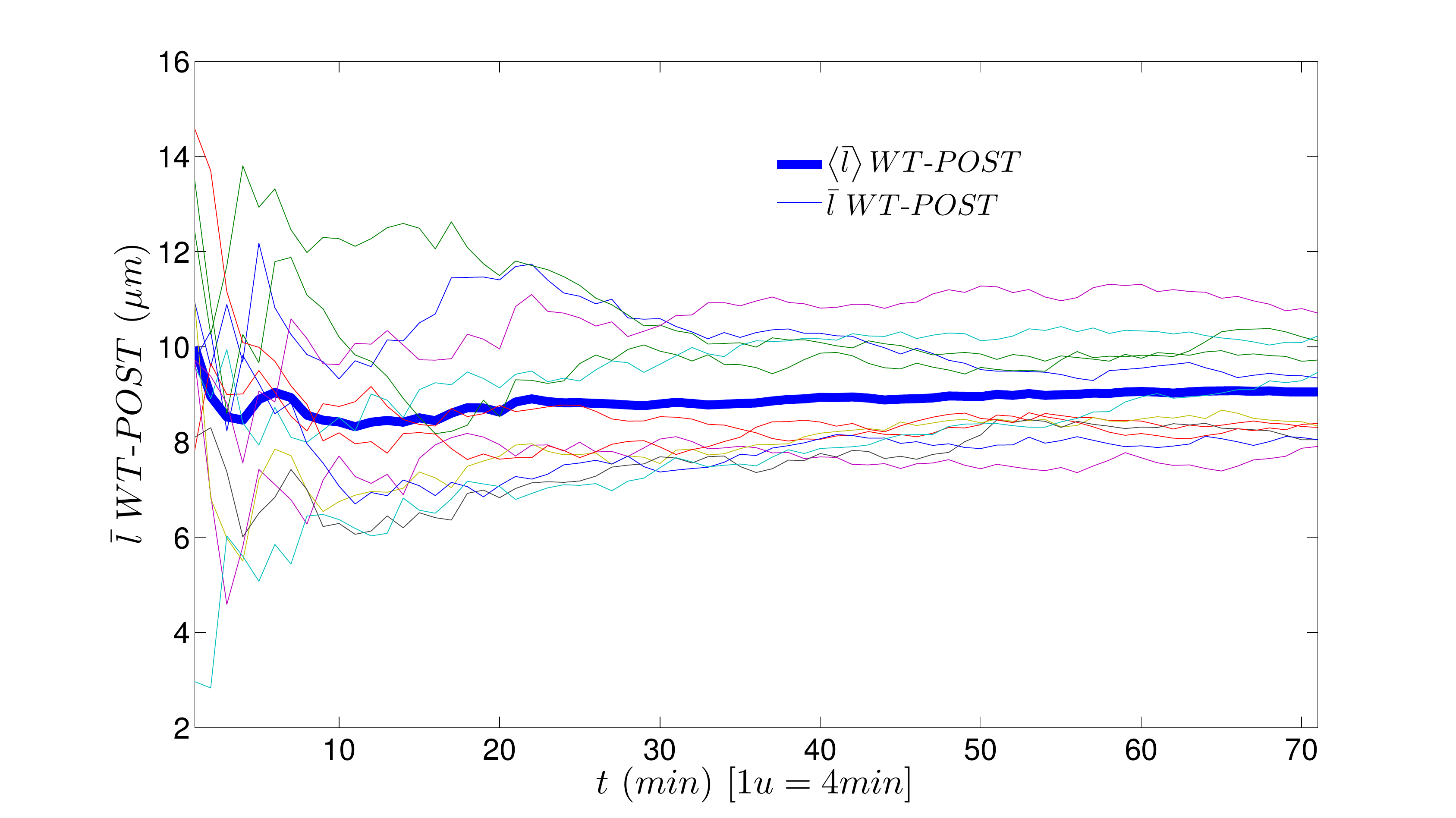}
\caption{Mean step length $l$ versus time for WT-PRE and WT-POST lymphocyte (lower and upper panel, respectively). Results for each splenocyte (thin curves) are compared with the resulting average over all splenocytes (thick curve). In both cases, $\overline{l}$ appears constant in time.}
\label{corr_temp_WT}
\end{figure}

\begin{figure} [!htbp]
\includegraphics[width=8.0cm]{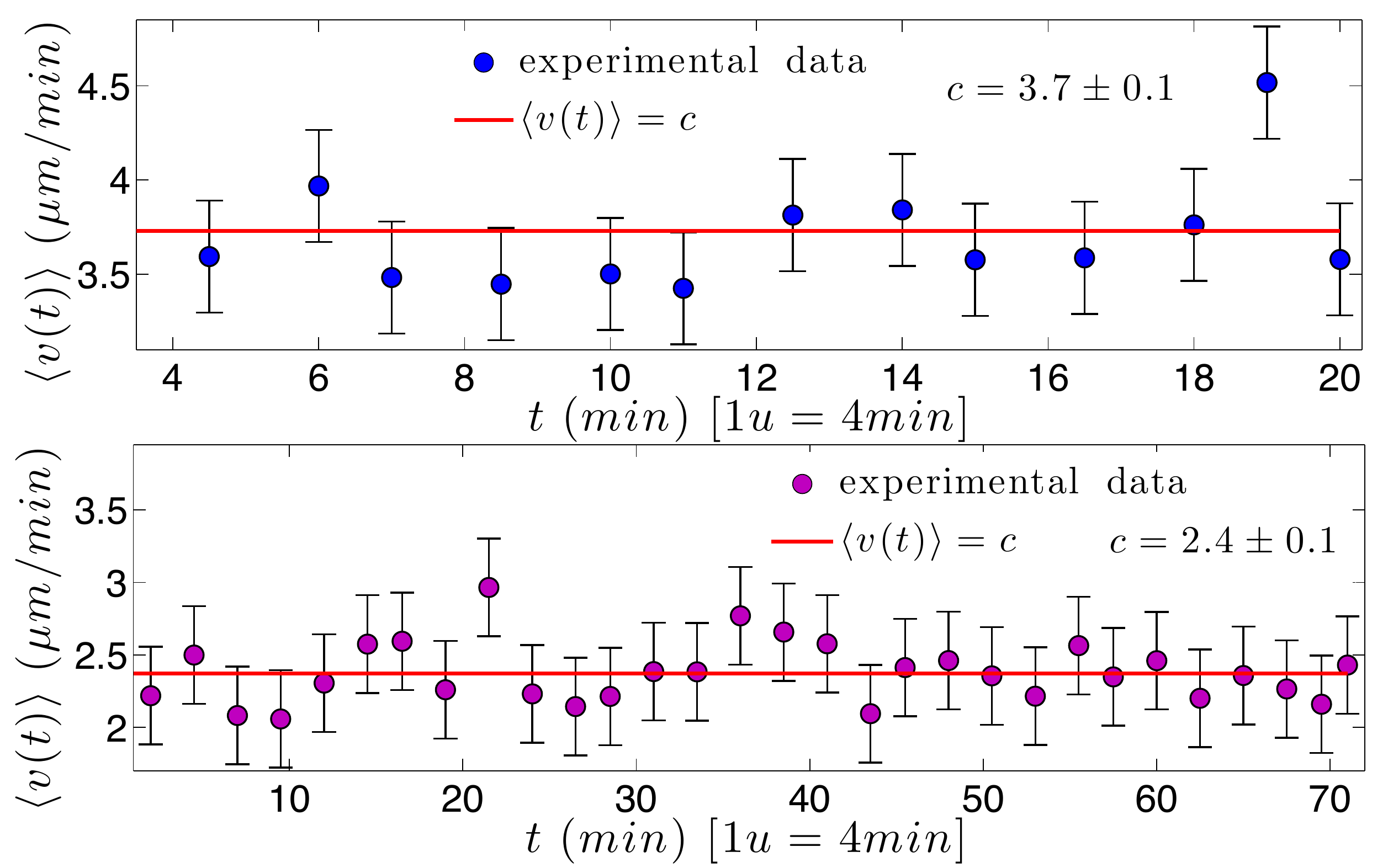}
\caption{Mean instantaneous speed of WT-PRE splenocytes (upper panel) and of WT-POST splenocytes (lower panel). Binned data ($\bullet$) with standard errors are best fitted by a constant line (solid curve). In both cases, no evident acceleration is observed, but, in the latter case, speed is lower.}
\label{v_WT}
\end{figure}

However, no significant temporal correlation among steps is evidenced since, for each splenocytes, the mean step $\overline{l}$ (see Eq.~\ref{passomedio}) turns out to be (approximately) constant in time for both WT-PRE and WT-POST splenocytes. Thus, no acceleration is observed in the process and the instantaneous speed is stable (see Fig.~\ref{corr_temp_WT} and \ref{v_WT}). Of note, the speed of the splenocytes decreases, once they have crossed the microchannels: while in the right channel it is $3.7 \pm 0.1$ $\mu m/min$, in the left channel it decreases down to $2.4 \pm 0.1$ $\mu m/min$. Again, we can notice how the behaviour of the splenocytes changes according to their proximity with the tumor.

\begin{figure} [!htbp]
\centering
\includegraphics[width=8cm]{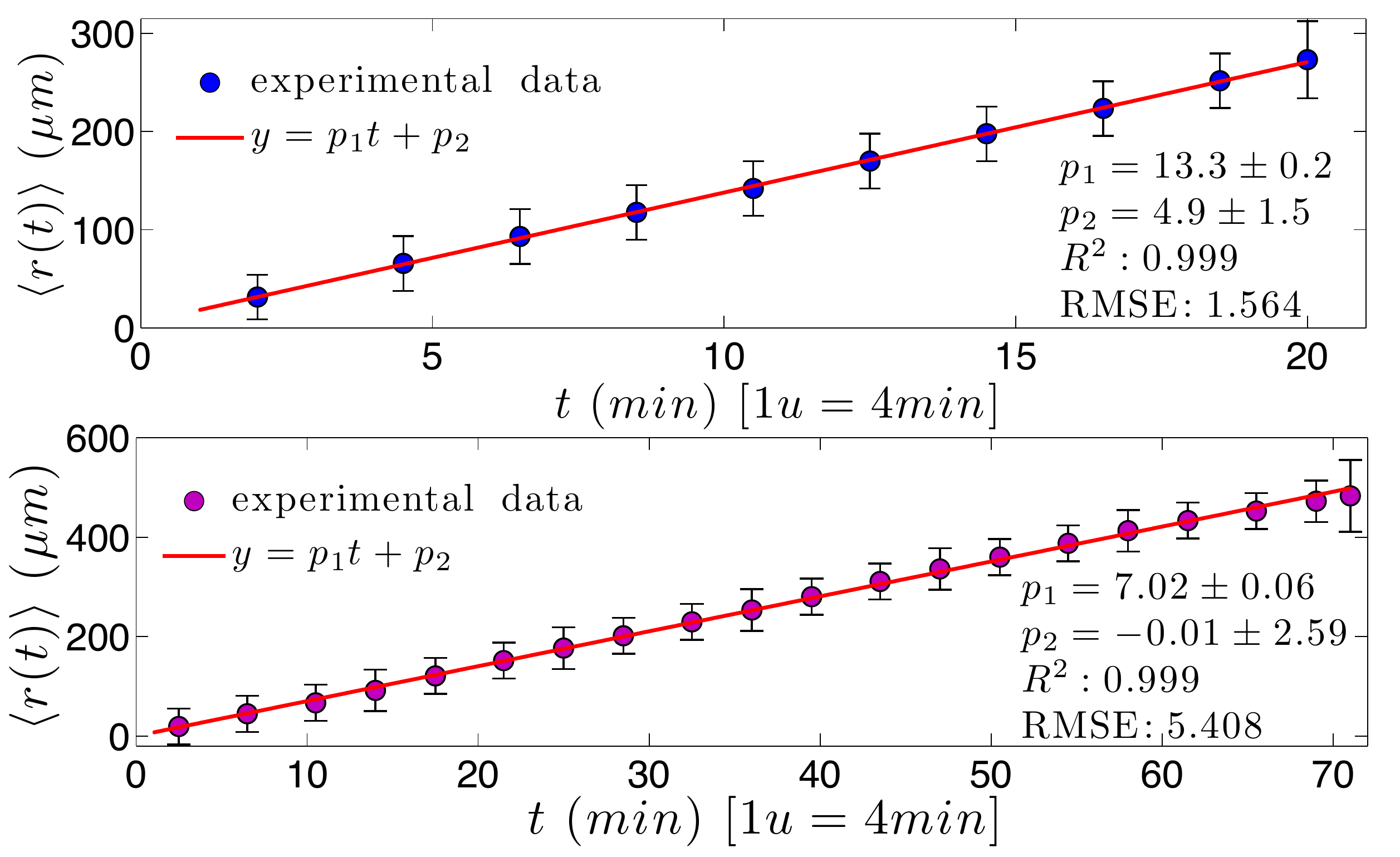}
\caption{$\left \langle r(t) \right \rangle$ versus $t$ for WT-PRE splenocytes (upper panel) and for WT-POST splenocytes (lower panel).  As expected, the average distance covered grows linearly with time. Binned data ($\bullet$) with standard errors are compared with best fit (solid line), whose coefficients are also reported.}
\label{r_WT}
\end{figure}

Focusing on the analysis of the macroscopic process, the most remarkable point is that the mean distance covered grows linearly with time (see Fig.$\ref{r_WT}$), for both WT-PRE and WT-POST splenocytes, as expected for a biased random walk, strongly supporting the evidence of a highly coordinate motion for the system as a whole. Notably, also in this case, it appears evident that WT-PRE splenocytes are faster than the WT-POST, since they cover a greater mean distance over time (see angular coefficients of the linear fit of $\langle r(t) \rangle$ in Table $\ref{ang_coeff_rWT}$).

Moreover, we checked that the linear behavior is also observed along both $x$ and $y$ directions of motion (of course, $\left \langle x(t) \right \rangle$ decreases and $\left \langle y(t) \right \rangle$ increases over time because the drift is directed along the negative $x$ and the positive $y$ axis).

In contrast to the isotropic unbiased random walk of KO splenocytes, here the mean square displacement is proportional to $t^2$ for large $t$ (see Fig.~\ref{ergodicity_WT}), so the signal propagates as a wave, as expected for a \textit{ballistic motion}.

\begin{figure} [!htbp]
\includegraphics[width=7cm, height=4cm]{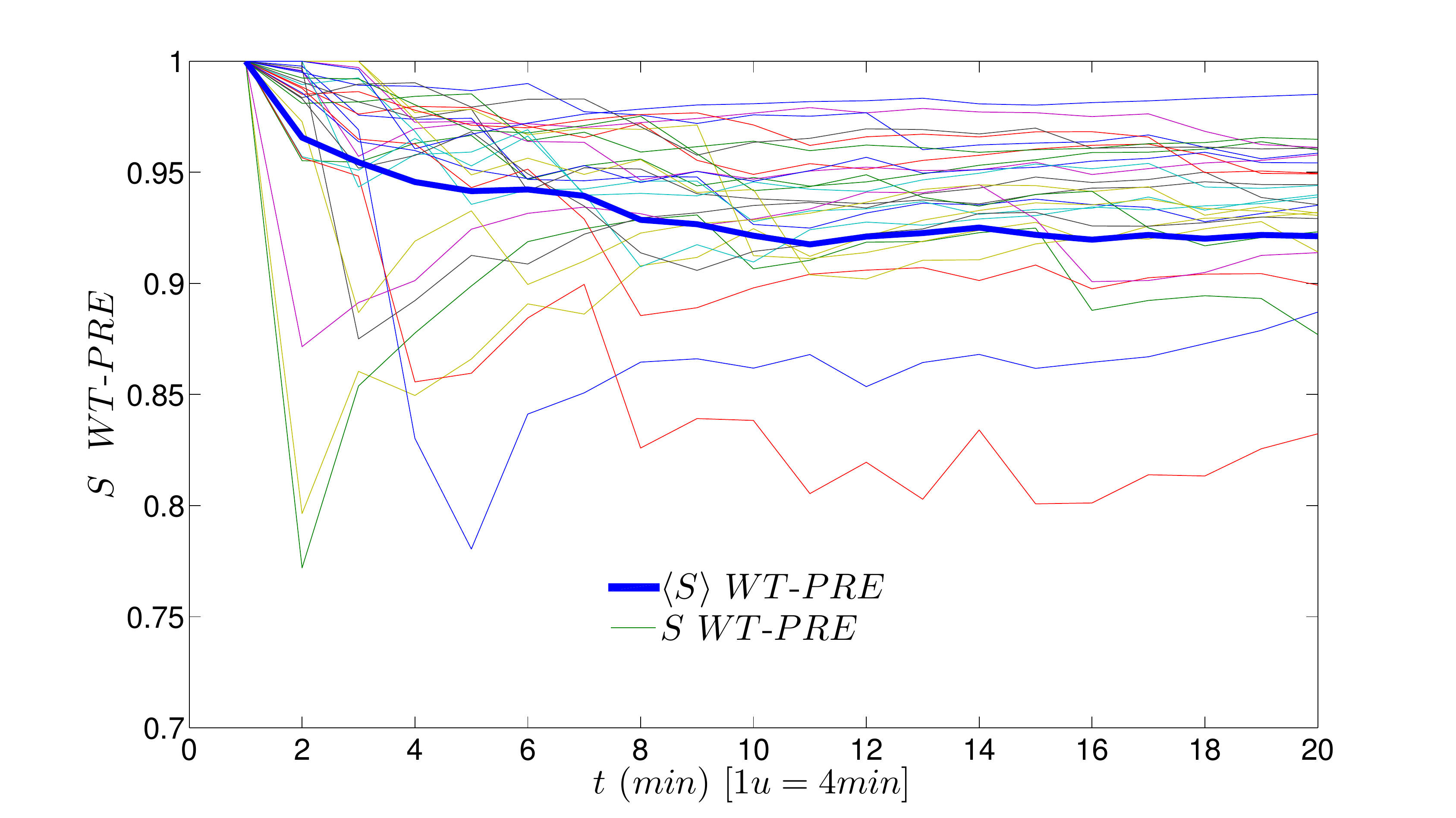}
\includegraphics[width=7cm, height=4cm]{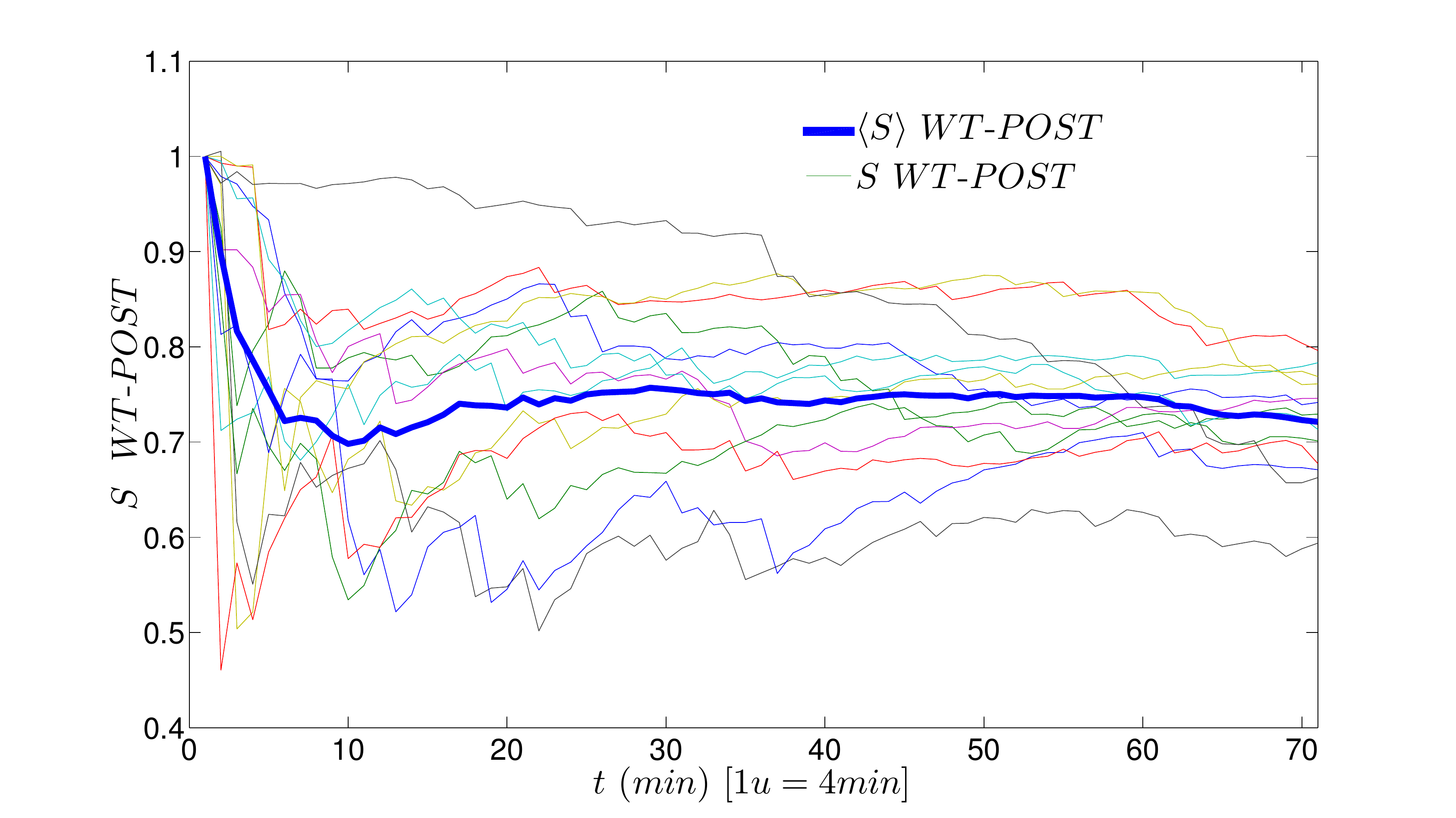}
\caption{Straightness index $S$ for WT-PRE splenocytes (upper panel) and for WT-POST splenocytes (lower panel); the thick lines represent the related averages over all splenocytes at each time $\left \langle S \right \rangle$. As expected for a random walk with bias, it takes values close to $1$.}
\label{S_WT}
\end{figure}
\begin{figure} [!htbp]
\includegraphics[width=8cm]{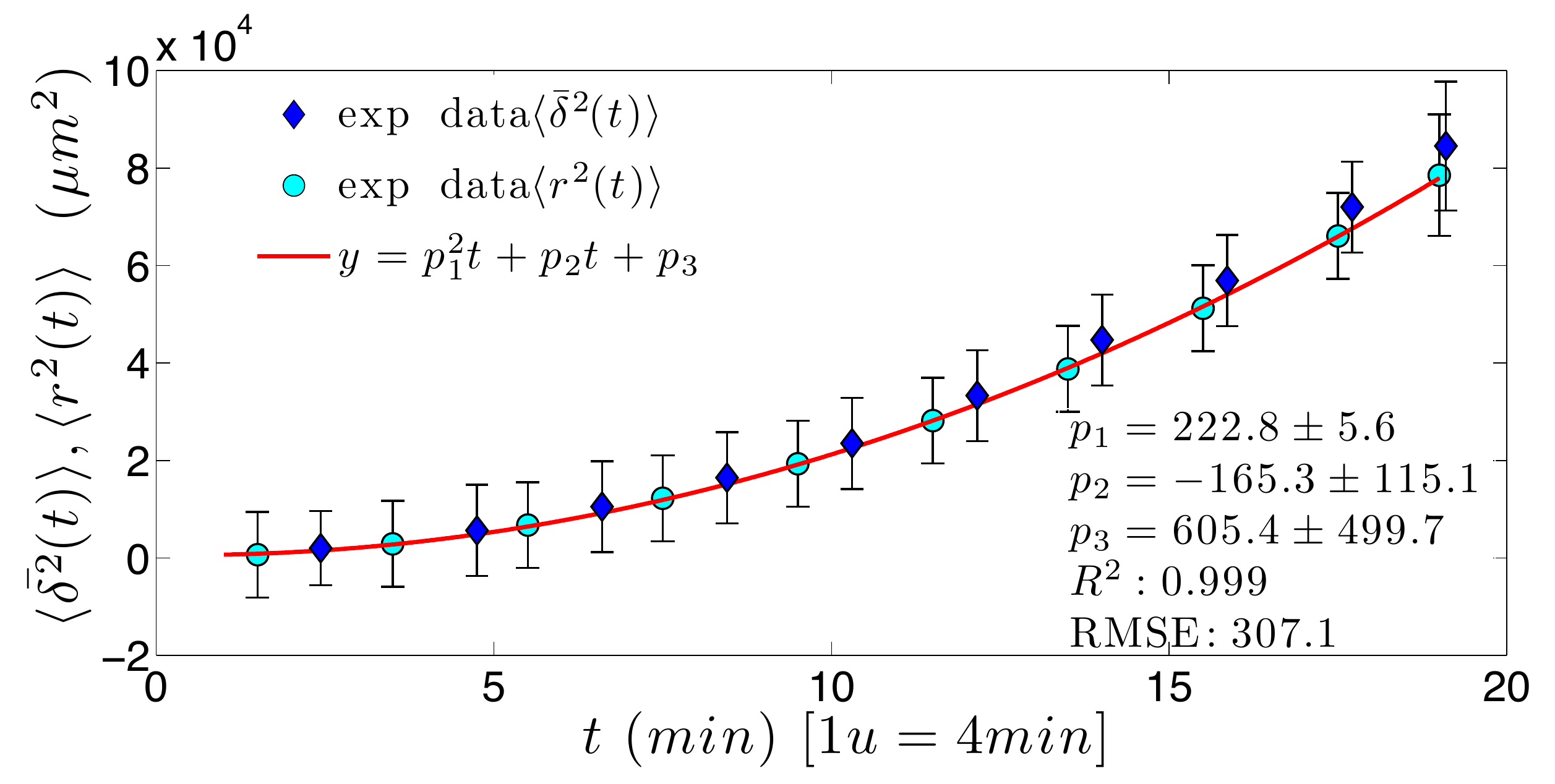}
\includegraphics[width=8cm]{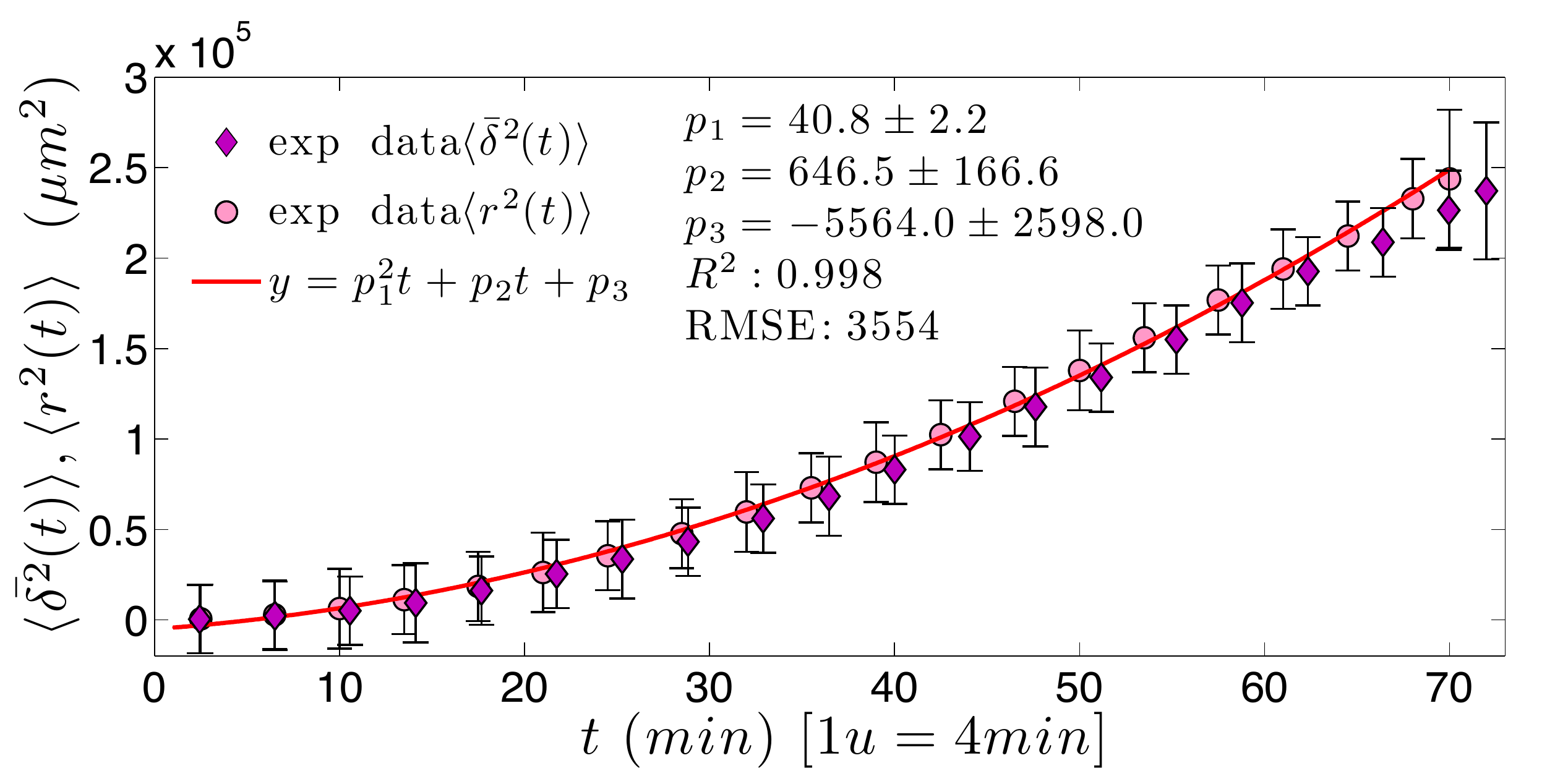}
\caption{$\langle \overline{\delta^2 (t)} \rangle$  ($\diamond$) and $\left \langle [r(t)] ^2 \right \rangle$ ($\circ$) vs $t$ for WT-PRE splenocytes (upper panel) and for WT-POST splenocytes (lower panel). Symbols represent experimental data with standard errors, while the solid line represents the best fits. Notice that, for both sets of lymphocytes, $\langle \overline{\delta^2 (t)} \rangle$ and $\left \langle [r(t)]^2 \right \rangle$ are nicely overlapped, both growing with the square of time.}
\label{ergodicity_WT}
\end{figure}

This picture of random walk with bias is also confirmed by the straightness index (see Fig.~\ref{S_WT}) whose mean value, for WT-PRE splenocytes ranges between $0.9$ and $1.0$, while for WT-POST, ranges between $0.7$ and $0.8$; this means that for WT-POST the motion is less straight, consistently with results discussed above.

\begin{table}[!htbp]
\begin{center}
\begin{tabular}{c | c }
\hline
\hline
Quantities & Angular coefficient \\
\hline
$\left \langle r(t) \right \rangle$ WT-PRE & 13.3 $\pm$ 0.1 \\
$\left \langle r(t) \right \rangle$ WT-POST & 7.0 $\pm$ 0.1 \\
\hline
\end{tabular}
\quad
\caption{Angular coefficients of linear fits for $\left \langle r(t) \right \rangle$ of WT-PRE and WT-POST splenocytes.}
\label{ang_coeff_rWT}
\end{center}
\end{table}

Finally, we successfully checked ergodicity also in WT splenocytes: as shown in Fig.~\ref{ergodicity_WT}, experimental data for $\langle \overline{\delta^2 (t)} \rangle$ and for $\left \langle r^2(t) \right \rangle$ are nicely overlapped and best-fitted by a power law with exponent approximately equal to $2$. Fit coefficients, reported in Tab.~\ref{deltaparameter_WT}, provide us with further important information about the role of melanoma cells on the motion of splenoctyes. Recalling that in an ergodic system $\overline{\delta^2} \sim 4Dt$, we obtain that the splenocytes that have passed the microchannels, move slower than those in the right channel. In other words, after crossing the microchannels, splenocytes tend to proceed slower.


\begin{table}[!htbp]
\begin{center}
\begin{tabular}{c | c | c}
\hline
\hline
Quantities & $p_1$ WT-PRE & $p_1$ WT-POST \\
\hline
$\left \langle \overline{\delta^2 (t)} \right \rangle$ & 222.8 $\pm$ 5.6 & 41.2 $\pm$ 1.5 \\
$\left \langle r^2(t) \right \rangle$ &  208 $\pm$ 8.9 &  40.7 $\pm$ 2.2 \\
\hline
\end{tabular}
\quad
\caption{Fit parameters of $\langle \overline{\delta^2 (t)} \rangle$ and $\left \langle [r(t)]^2 \right \rangle$ for WT-PRE and WT-POST splenocytes.}
\label{deltaparameter_WT}
\end{center}
\end{table}

\quad

\section{Conclusions}
Summarizing, in this work we collected and analyzed data on the motility of immune cells towards tumor cells by exploiting splenocytes deficient for the transcription factor IRF-8, in comparison to WT cells during co-culture with melanoma cells in a microfluidic device. In this experimental system, WT splenocytes are endowed with the ability to move towards the melanoma cells, so acting as a brake for cancer cell advancement.
\newline
By applying a quantitative description stemmed from stochastic process theory, we found deep differences in the migratory behavior of WT and IRF-8 KO splenocytes. Indeed, we found with remarkable accuracy that every single IRF-8 KO cell performs pure uncorrelated random walks without pointing to the target. Conversely, WT splenocytes are able to perform, singly, drifted random walks, which, collectively, collapse on a straight ballistic motion for the system as a whole, giving rise to an effectively high coordinate motion towards melanoma cells.
\newline
At a more detailed level of investigation, IRF-8 KO cells move rather uniformly since their step lengths are exponentially distributed with a characteristic step length $\lambda$ in agreement with literature, e.g. $\lambda \sim 4.5\mu m$. On the contrary, WT cells display a qualitatively broader motion, due to their step lengths along the direction of the melanoma log-normally distributed.
\newline
The resulting dynamics are in good agreement with models of in-vivo behavior of immune cells \cite{neoplasia,last}.
\newline
In conclusion, our data clearly evidence the value of Cell-on-Chip approaches as tools to easily perform ''under microscope'' experiments not only by simply tracking the migratory extent of cells inside the system, but also by applying specific mathematical models, such as those derived from the stochastic theory. In prospect, this will be of benefit for the development of the modern biology.

\quad


\section*{Acknowledgements}

Italian Minister of University and Research through the grant FIRB-RBFR08EKEV and through INdAM-GNFM support, Italian Association for Cancer Research (AIRC) project no. 11610 to LG, AIRC project no. 10720 and Italian Ministry of Health "Programma Integrato Oncologia" 2006 are acknowledged.
\newline
The authors are grateful to Giorgio Parisi fur useful discussions.




\footnotesize{
\bibliography{rsc} 
\bibliographystyle{rsc} 
}

\end{document}